\documentclass[aps,twocolumn,showpacs]{revtex4-1}
\usepackage{amssymb}
\usepackage{amsmath}
\usepackage{amscd}
\usepackage{graphicx}
\usepackage{subfigure}
\usepackage{float}
\usepackage{array}
\begin{document}

\title[Short Title]{Adiabatic approximation for three qubits ultrastrongly coupled to a harmonic oscillator}

%Three qubits couple to an harmonic oscillator by adiabatic approximation with ultrastrong coupling
%Three qubits couple to a harmonic oscillator adiabatically with ultrastrong coupling
%Three qubits couple to a harmonic oscillator with ultrastrong
%coupling by adiabatic approximation and the property of their ground
%analyze or behavior or Unified analytical treatments or

\author{Li-Tuo Shen}
%\email{lituoshen@gmail.com}
\author{Rong-Xin Chen}
%\author{Mei Lu}
\author{Huai-Zhi Wu}
\email{huaizhi.wu@fzu.edu.cn}
\author{Zhen-Biao Yang}
\email{zbyang@fzu.edu.cn}
%\author{Shi-Biao Zheng}
%\email{sbzheng11@163.com}

\affiliation{Lab of Quantum Optics, Department of Physics, Fuzhou
University, Fuzhou 350002, China}
%\begin{spacing}{2.0}

\begin{abstract}
We study the system involving mutual interaction between three
qubits and an oscillator within the ultrastrong coupling regime. We
apply adiabatic approximation approach to explore two extreme
regimes: (i) the oscillator's frequency is far larger than each
qubit's frequency and (ii) the qubit's frequency is far larger than
the oscillator's frequency, and analyze the energy-level spectrum
and the ground-state property of the qubit-oscillator system under
the conditions of various system parameters. For the energy-level
spectrum, we concentrate on studying the degeneracy in low energy
levels. For the ground state, we focus on its nonclassical
properties that are necessary for preparing the nonclassical states.
We show that the minimum qubit-oscillator coupling strength needed
for generating the nonclassical states of the Schr\"{o}dinger-cat
type in the oscillator is just one half of that in the Rabi model.
We find that the qubit-qubit entanglement in the ground state
vanishes if the qubit-oscillator coupling strength is strong enough,
for which the entropy of three qubits keeps larger than one. We also
observe the phase-transition-like behavior in the regime where the
qubit's frequency is far larger than the oscillator's frequency.
\end{abstract}

\pacs{42.50.Pq, 42.50.Nn, 42.50.Dv}
  \keywords{adiabatic approximation, three qubits, ultrastrongly coupled, harmonic oscillator}
\maketitle

\noindent
\section{Introduction}

The most fundamental model (known as the quantum Rabi model
\cite{PR-49-324-1936}) describing light-matter interactions exists
in the mutual coupling between a two-level system (or a qubit) and a
harmonic oscillator. Such a model is of great use in fields ranging
from quantum optics and quantum information \cite{BOOK-1} to
condensed matter physics \cite{AnnPhys-8-325-1959}. Typically,
experimental implementations of the kind of such a model have been
successful in many physical systems, such as QED cavities
\cite{RMP-73-565-2001}, ion traps \cite{RMP-75-281-2003},
nanomechanical resonators \cite{PRB-68-155311-2003}, solid systems
\cite{Nature-445-896-2007,Nature-445-515-2007}, etc.

In traditional cavity quantum electrodynamics (QED) systems
\cite{PRL-87-037902-2001,PRL-85-2392-2000}, the coupling between a
two-level atom and the cavity field is relatively weak as compared
to the atom's or field's frequency, thus the rotating-wave
approximation functions well and the Jaynes-Cummings (JC) model is
valid \cite{ProcIEEE-51-89-1963}. Recently, experimental
demonstrations associated with the qubit-oscillator system within
ultrastrong coupling regime have been reported
\cite{PRB-78-180502-2008,PRB-79-201303-2009,PRL-105-237001-2010,
PRL-105-196402-2010,PRL-106-196405-2011,Science-335-1323-2012,
PRL-108-163601-2012,PRB-86-045408-2012}, for which the
qubit-oscillator coupling rate reaches a significant fraction of the
qubit's or oscillator's frequency and the JC model is no longer
applicable, generating a plenty of fascinating quantum phenomena
\cite{PRA-87-023835-2013,NJP-13-073002-2011,PRL-109-193602-2012,PRA-74-033811-2006,
PRA-77-053808-2008,PRA-82-022119-2010,PRL-108-180401-2012}.

A natural generalization of the Rabi model is to involve $N$ qubits
simultaneously interacting with a common harmonic oscillator, i.e.,
the Dicke model
\cite{PR-93-99-1954,PRE-67-066203-2003,PRL-90-044101-2003}, in which
enormous interest has been devoted to investigating the dynamical
behavior in the thermodynamic limit $N\rightarrow\infty$, such as
superradiance phase transitions and entanglement properties
\cite{PRL-92-073602-2004,AP-76-360-1973,PRA-8-2517-1973,PRA-7-831-1973,PRA-84-033817-2011}.
Although many mathematics approaches, such as adiabatic
approximation \cite{PRL-99-173601-2007,PRB-72-195410-2005} and
transformation method \cite{EPJB-38-559-2004-2}, are presented to
analytically treat the Rabi model and two-qubit Dicke model within
the ultrastrong coupling regime
\cite{PRL-99-173601-2007,PRB-72-195410-2005,EPJB-38-559-2004-2,RPB-40-11326-1989,
RPB-42-6704-1990,EPL-86-54003-2009,PRA-80-033846-2009,
PRL-105-263603-2010,PRA-82-025802-2010,EPJD-66-1-2012,PRA-86-015803-2012,
PRA-85-043815-2012,PRA-86-023822-2012,PRL-107-100401-2011,PRB-75-054302-2007,
EPJD-59-473-2010,PRA-88-015802-2013,PRA-87-022124-2013,PRA-86-014303-2012,
JPA-46-415302-2013,JPA-46-335301-2013,PRA-81-042311-2010}, few
existing theoretical approaches can be directly applied to treat the
Dicke model involving a small number of qubits due to its
non-integrability in the Hilbert space
\cite{EPL-90-54001-2010,JPB-46-224016-2013}, which is believed to
possess richer dynamics properties and more potential applications
in quantum information processing than that in the Rabi model
\cite{PRL-107-190402-2011}. To our knowledge, the Dicke model with
three or a little bit more qubits has not been extensively
investigated. Recently, Tsomokos \emph{et al.} analyzed a similar
model and derived a low-energy Hamiltonian for the three qubits by
adiabatically eliminating the resonator \cite{NJP-10-113020-2008}.
However, their approach does not apply in the ultrastrong-coupling
regime, where the state of the resonator is strongly dependent on
the state of the qubits. Braak \cite{arXiv-1304-2529-2013} has
analytically obtained the spectrum of the Dicke model with three
qubits based on the formal but complicated solutions in the Bargmann
space. We also achieved the approximately analytical ground state in
the Dicke model with three qubits by using the transformation method
\cite{arXiv-1305-1226-2013}. Both of the studies
\cite{arXiv-1304-2529-2013,arXiv-1305-1226-2013} mainly discuss the
near-resonant mechanism where the oscillator's frequency is close to
the qubit's frequency.

Differing in such previous studies
\cite{arXiv-1304-2529-2013,arXiv-1305-1226-2013}, we explore here
two further regimes for the system of three qubits ultrastrongly
coupled to an oscillator, i.e., the oscillator's frequency is far
larger than each qubit's frequency (say, a high-frequency
oscillator) and each qubit's frequency is far larger than the
oscillator's frequency (say, three high-frequency qubits). We use
the approach of adiabatic approximation in which the bias of each
qubit is distinctly involved, and focus on analyzing their
energy-level spectra and ground-state properties with the choice of
various system parameters. For the energy-level spectrum, we
concentrate on studying the degeneracy in low energy levels. For the
ground state, we obtain its nonclassical properties that are
necessary for preparing the nonclassical states, including the
squeezed state of the oscillator, the Schr\"{o}dinger-cat state of
the oscillator, the qubit-oscillator entangled state, and the
qubit-qubit entangled state. Different from the Rabi model
\cite{PRA-81-042311-2010}, the qubit-oscillator coupling strength
needed here for generating nonclassical states of the
Schr\"{o}dinger-cat type in the oscillator is much smaller.
Particularly, we find that the qubit-qubit entanglement in the
ground state vanishes if the qubit-oscillator coupling strength is
strong enough, in which the entropy of three qubits keeps larger
than one. Interestingly, we observe the phase-transition-like
behaviors in the regime where each qubit's frequency is far larger
than the oscillator's frequency \cite{PRA-81-042311-2010-2}, which
is very different from the phase transition found in the resonant
regime
\cite{PRL-92-073602-2004,AP-76-360-1973,PRA-8-2517-1973,PRA-7-831-1973,PRA-84-033817-2011}.
Possible experiment realization of our generalized Dicke model with
three high-frequency qubits or a high-frequency oscillator can be
expected in the superconducting experiment \cite{Nature-6-772-2010},
where the flexibility of controlling kinds of system parameters
allows one to unearth rich dynamics properties of the system within
the ultrastrong coupling regime.

\section{System Hamiltonian}

\begin{figure}
\center
  \includegraphics[width=0.6\columnwidth]{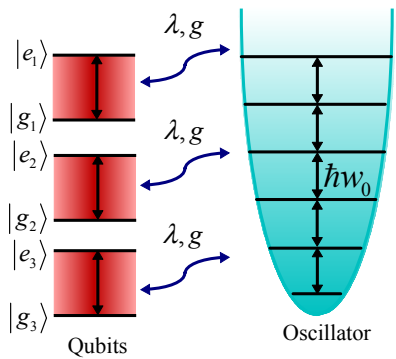} \caption{(Color
  online) Schematic of the system with three identical qubits coupled to
  a harmonic
  oscillator. The $j$th ($j=1,2,3$) qubit with one ground ($|g_j\rangle$) and
  one excited states ($|e_j\rangle$) is coupled to the
  oscillator with frequency $w_0$, where the qubit-oscillator coupling
  strength is denoted by $g$ or $\lambda$.
  }\label{Fig.3.}
\end{figure}

The quantum system we consider here is that three identical qubits
couple to a common harmonic oscillator, as shown in Fig. 1. The
Hamiltonian of this qubit-oscillator system is written as :
\begin{eqnarray}\label{e1}
\hat{H}&=&\sum_{j=1}^{3}\hat{H}_{q_{j}} + \hat{H}_{ho} +
\hat{H}_{int},
\end{eqnarray}
where
\begin{eqnarray}\label{e2}
\hat{H}_{q_{j}}&=&-\frac{\Delta}{2}\hat{\sigma}_{x_{j}}-\frac{\epsilon}{2}\hat{\sigma}_{z_{j}},\cr\cr
\hat{H}_{ho}&=&\frac{\hat{p}^2}{2m} +
\frac{1}{2}mw_{0}^2\hat{x}^2,\cr\cr
\hat{H}_{int}&=&g\hat{x}(\hat{\sigma}_{z_{1}}+\hat{\sigma}_{z_{2}}+\hat{\sigma}_{z_{3}}),
\end{eqnarray}
$\hat{\sigma}_{x_{j}}$ and $\hat{\sigma}_{z_{j}}$ are the Pauli
matrices (with $\hat{\sigma}_{z_{j}}|e_{j}\rangle$ =
$|e_{j}\rangle$, $\hat{\sigma}_{z_{j}}|g_{j}\rangle$ =
$-|g_{j}\rangle$, $\hat{\sigma}_{x_{j}}|e_{j}\rangle$ =
$|g_{j}\rangle$, and $\hat{\sigma}_{x_{j}}|g_{j}\rangle$ =
$|e_{j}\rangle$) of the $j$th qubit, where $|e_{j}\rangle$ and
$|g_{j}\rangle$ are respectively the excited and ground states of
the $j$th qubit, and $\hat{x}$ and $\hat{p}$ are respectively the
position and momentum operators of the harmonic oscillator. $\Delta$
and $\epsilon$ are the gap and bias characterizing each qubit. $m$
is the effective mass of the oscillator with frequency $w_{0}$, and
$g$ is the coupling strength between each qubit and the oscillator.
The parameters $\Delta$, $\epsilon$, and $g$ are assumed to be
positive and real for convenience in the paper.

We can also use the creation ($\hat{a}^{\dagger}$) and annihilation
($\hat{a}$) operators to express the oscillator's Hamiltonian for
convenience :
\begin{eqnarray}\label{e3}
\hat{H}_{ho}&=&\hbar w_0\hat{a}^{\dagger}\hat{a}, \cr\cr
 \hat{H}_{int}&=&\lambda(\hat{a}^{\dagger}+\hat{a})(\hat{\sigma}_{z_{1}}+\hat{\sigma}_{z_{2}}+\hat{\sigma}_{z_{3}}),
\end{eqnarray}
where the zero-point energy $\frac{1}{2}\hbar w_{0}$ in
$\hat{H}_{ho}$ is omitted, and
\begin{eqnarray}\label{e4}
\hat{a}^{\dagger}&=&\hat{X}-i\hat{P}=\sqrt{\frac{mw_{0}}{2\hbar}}\hat{x}-i\frac{\hat{p}}{\sqrt{2\hbar
mw_{0}}},\cr \cr
\hat{a}&=&\hat{X}+i\hat{P}=\sqrt{\frac{mw_{0}}{2\hbar}}\hat{x}+i\frac{\hat{p}}{\sqrt{2\hbar
mw_{0}}},\cr \lambda&=&\sqrt{\frac{\hbar}{2mw_{0}}}g.
\end{eqnarray}

In the basis $\{|e_{j}\rangle, |g_{j}\rangle\}$ of the $j$th qubit,
the eigenenergies of $\hat{H}_{q_{j}}$ are $\pm E_{q}/2$, where
$E_{q}=\sqrt{\Delta^2+\epsilon^2}$ is the eigenenergy splitting of
the bare qubit, and the energetic excited and ground states are
respectively represented by $|\uparrow_j\rangle$ and
$|\downarrow_j\rangle$. The angle $\theta$ is introduced here and
defined as $\tan\theta=\epsilon/\Delta$, which denotes the relative
size between $\hat{\sigma}_{x_{j}}$ and $\hat{\sigma}_{z_{j}}$ and
will be used later. The eigenstates of $\hat{H}_{ho}$ are supposed
to be the $n$-photon Fock state $|n\rangle$ ($n=0, 1, 2, ...$) with
the corresponding eigenenergy $n\hbar w_{0}$ in the bare oscillator.

Since three qubits are equivalent, the system Hamiltonian $\hat{H}$
is rotationally invariant which leads to a splitting of the
eight-dimensional Hilbert space of the qubits into three irreducible
components and ensuing degeneracies, on the basis of the formula
$\frac{1}{2}\otimes\frac{1}{2}\otimes\frac{1}{2}=\frac{1}{2}\oplus\frac{1}{2}\oplus\frac{3}{2}$
\cite{arXiv-1304-2529-2013}. The three-qubit Dicke model is
mathematically equivalent to two single-qubit Rabi models and a
spin-$3/2$ system. However, this symmetry characteristic does not
result in integrability because a spin-$3/2$ system has a
four-dimensional state space and the nontrivial eigenstates cannot
be labelled by a quantum number \cite{arXiv-1304-2529-2013}. We
remark that since the merit of our adiabatic-approximation method is
to give the intuitive insight into the physics, the collective
operators and nontrivial vectors for the qubit states are not used
and the symmetry characteristic is included in the degeneracy
analysis of eigenstates and eigenenergies after using adiabatic
approximation in two following situations.

\section{Adiabatic approximation for two extreme regimes}

In the following section, we focus on exploring the qubit-oscillator
system in two extreme regimes that can be analytically treated
through the approach of adiabatic approximation.

\subsection{High-frequency oscillator}

The first extreme situation we study is that the oscillator's
frequency is far larger than the qubit's eigenenergy splitting
(i.e., $\hbar w_0>>E_{q}$) as well as the coupling strength (i.e.,
$\hbar w_0>>g,\lambda$), in which the oscillator approximately keeps
in its energy eigenstate and adiabatically follows the changes
induced by three qubits' states.

Based on the idea of adiabatic approximation
\cite{PRL-99-173601-2007,PRA-81-042311-2010}, we assume that the
$j$th qubit has a well defined value of $\hat{\sigma}_{z_{j}}$,
i.e., $\hat{\sigma}_{z_{1}}=\pm 1$, $\hat{\sigma}_{z_{2}}=\pm 1$,
and $\hat{\sigma}_{z_{3}}=\pm 1$. For simplicity, we define four
collective-state symbols for three qubits: $|A_{+3}\rangle$
($|A_{-3}\rangle$) represents that three qubits are all in their
excited states $|e_{1}e_{2}e_{3}\rangle$ (ground states
$|g_{1}g_{2}g_{3}\rangle$); $|A_{+1}\rangle$ represents that one of
three qubits is in its ground state and the other two qubits are
both in the excited states, i.e., $|e_{1}e_{2}g_{3}\rangle$ or
$|e_{1}g_{2}e_{3}\rangle$ or $|g_{1}e_{2}e_{3}\rangle$;
$|A_{-1}\rangle$ represents that one of three qubits is in its
excited state and the other two qubits are both in the ground
states, i.e., $|e_{1}g_{2}g_{3}\rangle$ or $|g_{1}e_{2}g_{3}\rangle$
or $|g_{1}g_{2}e_{3}\rangle$. Therefore, when the qubits are in the
state $|A_{\pm1}\rangle$, the effective Hamiltonian for the
oscillator is :
\begin{eqnarray}\label{e5}
\hat{H}_{ho,|A_{\pm1}\rangle}&=&\hbar w_{0}\hat{a}^{\dagger}\hat{a}
\pm \lambda(\hat{a}^{\dagger}+\hat{a}),
\end{eqnarray}
which is the same as that in the system with one qubit and one
oscillator \cite{PRA-81-042311-2010}.

When the qubits are in the state $|A_{\pm3}\rangle$, the effective
Hamiltonian for the oscillator is :
\begin{eqnarray}\label{e6}
\hat{H}_{ho,|A_{\pm3}\rangle}&=&\hbar w_{0}\hat{a}^{\dagger}\hat{a}
\pm 3\lambda(\hat{a}^{\dagger}+\hat{a}),
\end{eqnarray}
which is interpreted as the original oscillator Hamiltonian with the
term of a larger force applied, as compared to that applied in the
system of one or two qubits interacting with an oscillator
\cite{PRA-81-042311-2010,PLA-376-2977-2012}. The Hamiltonian in Eqs.
(\ref{e5}) and (\ref{e6}) can be respectively transformed into Eqs.
(7) and (8) :
\begin{eqnarray}\label{e7-8}
\hat{H}_{ho,|A_{\pm1}\rangle}&=&
\hbar w_{0}\hat{a}^{\dagger}_{\pm1}\hat{a}_{\pm1}-\frac{\lambda^2}{\hbar w_0},\\
\hat{H}_{ho,|A_{\pm3}\rangle}&=& \hbar
w_{0}\hat{a}^{\dagger}_{\pm3}\hat{a}_{\pm3}-\frac{9\lambda^2}{\hbar
w_0},
\end{eqnarray}
with
\begin{eqnarray}\label{e9-10}
\hat{a}_{\pm1}&=&\hat{a} \pm \frac{\lambda}{\hbar w_0},\\
\hat{a}_{\pm3}&=&\hat{a} \pm \frac{3\lambda}{\hbar w_0}.
\end{eqnarray}
The eigenstates of $\hat{H}_{ho,|A_{\pm1}\rangle}$ and
$\hat{H}_{ho,|A_{\pm3}\rangle}$ are the displaced Fock states
\cite{PRA-46-4138-1992}:
\begin{eqnarray}\label{e11-12}
|n_{\pm 1}\rangle&=&e^{\mp \frac{\lambda}{\hbar w_{0}}(\hat{a}^{\dagger}-\hat{a})}|n\rangle,\\
|n_{\pm 3}\rangle&=&e^{\mp \frac{3\lambda}{\hbar
w_{0}}(\hat{a}^{\dagger}-\hat{a})}|n\rangle,
\end{eqnarray}
with the corresponding eigenenergies :
\begin{eqnarray}\label{e13-14}
E_{o,\pm1}&=&n\hbar w_{0}-\frac{\lambda^2}{\hbar w_0},\\
E_{o,\pm3}&=&n\hbar w_{0}-\frac{9\lambda^2}{\hbar w_0}.
\end{eqnarray}
The above results can also be intuitively interpreted as the
corresponding harmonic oscillator potentials in the
position-momentum picture, as plotted in Fig. 2.
\begin{figure}
\center
  \includegraphics[width=1\columnwidth]{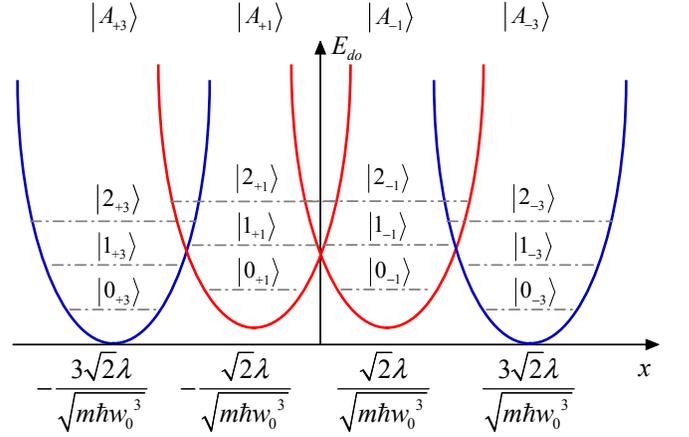} \caption{(Color
  online) Schematic of four displaced oscillators. The horizontal and
  vertical axises represent the position and displaced oscillator's
  eigenenergy $E_{do}$, respectively. Four displaced oscillators are shifted
  to the left or right from the equilibrium position with a specific constant,
  where the shift direction is determined by the
  state of three qubits. The eigenstates (plotted with $n$ no more
  than $2$) that have the same value of $n$ are degenerate for the states
  $|A_{\pm1}\rangle$ (or $|A_{\pm3}\rangle$), and have the symmetry divided by the
  origin point in horizontal axis.
  }\label{Fig.2.}
\end{figure}
The eigenstates of the oscillator slightly rely on the states of
three qubits, among which there exists special orthogonality, i.e.,
$\langle m_{-3}|n_{-3}\rangle$ $=\delta_{mn}$, $\langle
m_{-1}|n_{-1}\rangle$ $=\delta_{mn}$, $\langle m_{+1}|n_{+1}\rangle$
$=\delta_{mn}$, $\langle m_{+3}|n_{+3}\rangle$ $=\delta_{mn}$, and
\begin{eqnarray}\label{e15}
&&\langle m_{+3}|n_{+1}\rangle=\langle m_{+1}|n_{-1}\rangle=\langle
m_{-1}|n_{-3}\rangle\cr\cr&&=
  \left\{
   \begin{array}{l}
   \sqrt{\frac{m!}{n!}} e^{-\frac{2\lambda^2}{(\hbar w_{0})^{2}}}(-\frac{2\lambda}{\hbar w_{0}})^{n-m} L_{m}^{n-m}\big[(\frac{2\lambda}{\hbar w_{0}})^{2}\big],n\geq m, \cr\cr
   \sqrt{\frac{n!}{m!}} e^{-\frac{2\lambda^2}{(\hbar w_{0})^{2}}}(\frac{2\lambda}{\hbar w_{0}})^{m-n} L_{n}^{m-n}\big[(\frac{2\lambda}{\hbar w_{0}})^{2}\big],m> n,
   \end{array}
  \right.\cr\cr
  &&\langle m_{+3}|n_{-1}\rangle=\langle m_{+1}|n_{-3}\rangle\cr\cr&&=
  \left\{
   \begin{array}{l}
   \sqrt{\frac{m!}{n!}} e^{-\frac{8\lambda^2}{(\hbar w_{0})^{2}}}(-\frac{4\lambda}{\hbar w_{0}})^{n-m} L_{m}^{n-m}\big[(\frac{4\lambda}{\hbar w_{0}})^{2}\big],n\geq m, \cr\cr
   \sqrt{\frac{n!}{m!}} e^{-\frac{8\lambda^2}{(\hbar w_{0})^{2}}}(\frac{4\lambda}{\hbar w_{0}})^{m-n} L_{n}^{m-n}\big[(\frac{4\lambda}{\hbar w_{0}})^{2}\big],m> n,
   \end{array}
  \right.\cr\cr
  &&\langle m_{+3}|n_{-3}\rangle\cr\cr&&=
  \left\{
   \begin{array}{l}
   \sqrt{\frac{m!}{n!}} e^{-\frac{18\lambda^2}{(\hbar w_{0})^{2}}}(-\frac{6\lambda}{\hbar w_{0}})^{n-m} L_{m}^{n-m}\big[(\frac{6\lambda}{\hbar w_{0}})^{2}\big],n\geq m, \cr\cr
   \sqrt{\frac{n!}{m!}} e^{-\frac{18\lambda^2}{(\hbar w_{0})^{2}}}(\frac{6\lambda}{\hbar w_{0}})^{m-n} L_{n}^{m-n}\big[(\frac{6\lambda}{\hbar w_{0}})^{2}\big],m> n,
   \end{array}
  \right.\cr&&
\end{eqnarray}
where $\delta_{mn}$ is the delta function, and $L_{m}^{n-m}$ and
$L_{n}^{m-n}$ are the associated Laguerre polynomials.

Based on the eigenstates of displaced oscillator, we now turn to the
qubits of the system. For a special value $n$ in the displaced
oscillator's state, we obtain an effective Hamiltonian for three
qubits. There are eight qubit states for each value of $n$, then the
effective Hamiltonian $\hat{H}_{q,eff}$ is a $8\times8$ matrix
expanded in the space $\Gamma\equiv\{$ $|\phi_{1}\rangle$,
$|\phi_{2}\rangle$, $|\phi_{3}\rangle$, $|\phi_{4}\rangle$,
$|\phi_{5}\rangle$, $|\phi_{6}\rangle$, $|\phi_{7}\rangle$,
$|\phi_{8}\rangle$ $\}$ $\equiv$ $\{$ $|eee\rangle|n_{+3}\rangle$,
$|eeg\rangle|n_{+1}\rangle$, $|ege\rangle|n_{+1}\rangle$,
$|gee\rangle|n_{+1}\rangle$, $|egg\rangle|n_{-1}\rangle$,
$|geg\rangle|n_{-1}\rangle$, $|gge\rangle|n_{-1}\rangle$,
$|ggg\rangle|n_{-3}\rangle$ $\}$. By diagonalizing $\hat{H}_{q,eff}$
in $\Gamma$, we obtain its eigenstates $|\phi_{qj}\rangle$ and
eigenenergies $E_{qj}$ ($j=1,2,...,8$):
\begin{eqnarray}\label{e16}
E_{q1}&=&E_{q2}=E_{q3}=E_{q,+1}, \cr
|\varphi_{q1}\rangle&=&|\Phi_1(+)\rangle,|\varphi_{q2}\rangle=|\Phi_2(+)\rangle,|\varphi_{q3}\rangle=|\Phi_3(+)\rangle,\cr\cr
E_{q4}&=&E_{q5}=E_{q6}=E_{q,-1}, \cr
|\varphi_{q4}\rangle&=&|\Phi_1(-)\rangle,|\varphi_{q5}\rangle=|\Phi_2(-)\rangle,|\varphi_{q6}\rangle=|\Phi_3(-)\rangle,\cr\cr
E_{q7}&=&E_{q,+3}, \cr
|\varphi_{q7}\rangle&=&|\Phi_4(+)\rangle,\cr\cr E_{q8}&=&E_{q,-3},
\cr |\varphi_{q8}\rangle&=&|\Phi_4(-)\rangle,
\end{eqnarray}
where
\begin{eqnarray}\label{e17}
E_{q,\pm1}&=&\pm\frac{\Delta}{2}\sqrt{l^2+\tan^2\theta},\cr\cr
E_{q,\pm3}&=&\pm\frac{3\Delta}{2}\sqrt{l^2+\tan^2\theta},\cr\cr
 |\Phi_1(\pm)\rangle&=&B(\mp)\big(\pm|\phi_3\rangle\mp|\phi_4\rangle-|\phi_5\rangle+|\phi_6\rangle\big),\cr
 |\Phi_2(\pm)\rangle&=&B(\mp)\big(\pm|\phi_3\rangle\mp|\phi_4\rangle-|\phi_5\rangle+|\phi_7\rangle\big),\cr
 |\Phi_3(\pm)\rangle&=&B(\mp)\big[|\phi_1\rangle-\frac{2\tan\theta}{l}(|\phi_2\rangle+|\phi_3\rangle)\big]\cr&&
 +(\frac{\pm2B(\mp)\tan\theta
 }{l}-1)|\phi_4\rangle\cr&&\mp(B(\mp)\mp\frac{2\tan\theta}{l})|\phi_5\rangle+|\phi_8\rangle,\cr\cr
 |\Phi_4(\pm)\rangle&=&\frac{\mp(l^2+4\tan^2\theta)B(\mp)+2l\tan\theta}{l^2}|\phi_1\rangle\cr\cr&&
 +(\frac{\mp2B(\mp)\tan\theta
 }{l}+1)(|\phi_2\rangle+|\phi_3\rangle+|\phi_4\rangle)\cr&&
 \mp B(\mp)(|\phi_5\rangle+|\phi_6\rangle+|\phi_7\rangle)+|\phi_8\rangle,\cr\cr
 B(\pm)&=&\frac{\sqrt{l^2+\tan^2\theta}\pm\tan\theta}{l},\cr\cr
 l&=&e^{-2\lambda^2/(\hbar w_{0})^{2}}L_{n}\big[(\frac{2\lambda}{\hbar w_0})^2\big].
\end{eqnarray}
Note that if three qubits are in the state $|A_{+1}\rangle$ or
$|A_{-1}\rangle$, there are three degenerate eigenstates for
$\hat{H}_{q,eff}$ in any given value $n$ of the oscillator. We
emphasize that the results discussed here still hold even when
$\lambda\geq \hbar w_0$ \cite{PRA-81-042311-2010}, under which the
oscillator can be adjusted adiabatically to the slow processes
governed by the gaps of non-degenerate eigenenergies. Especially,
both the eigenstates and eigenenergies of $\hat{H}_{q,eff}$ depend
on the number of photons $n$ in the oscillator, and the gap between
any two non-degenerate eigenenergies decreases as a Gaussian
function with increasing $\lambda/(\hbar w_0)$ which becomes a
constant when $\lambda/(\hbar w_0)$ is large enough, indicating that
eight eigenstates in $\Gamma$ decouple with each other.

\subsection{High-frequency qubits}

The second extreme situation we study is that the eigenenergy
splitting $E_{q}$ of each qubit is far larger than $\hbar w_0$ and
$\lambda$, in which three qubits approximately remain in their
energy eigenstates and adiabatically follow the changes induced by
the slow oscillator. Similar to the adiabatic approximation
discussed in Sec. IIIA, we assume that the oscillator has a
well-defined value $x$ of the position operator and deal with the
effective Hamiltonian of three qubits \cite{PRA-81-042311-2010}:
\begin{eqnarray}\label{e18}
\hat{H}_{hq,eff}&=&\sum_{j=1}^{3}(-\frac{\Delta}{2}\hat{\sigma}_{x_{j}}-\frac{\epsilon}{2}\hat{\sigma}_{z_{j}})+gx(\hat{\sigma}_{z_{1}}+\hat{\sigma}_{z_{2}}+\hat{\sigma}_{z_{3}}),\cr&&
\end{eqnarray}
for which we obtain the eigenenergies as follows :
\begin{eqnarray}\label{e19}
E_{hq,-3}&=&-3\sqrt{\frac{\Delta^2}{4}+(gx-\frac{\epsilon}{2})^2},\cr
E_{hq,-1}&=&-\sqrt{\frac{\Delta^2}{4}+(gx-\frac{\epsilon}{2})^2} \ \
(\mathrm{triple \ degenerate}),\cr
E_{hq,+1}&=&\sqrt{\frac{\Delta^2}{4}+(gx-\frac{\epsilon}{2})^2} \ \
(\mathrm{triple \ degenerate}),\cr
E_{hq,+3}&=&3\sqrt{\frac{\Delta^2}{4}+(gx-\frac{\epsilon}{2})^2}.
\end{eqnarray}
Since the eigenenergies of three high-frequency qubits depend on the
position $x$ of the oscillator, we now turn to the oscillator's
effective potentials to gain more transparent physics. Acquiring new
contributions from three qubits, the oscillator's effective
potentials are obtained :
\begin{eqnarray}\label{e20-21}
V_{eff,\pm1}&=&\frac{1}{2}mw_{0}^{2}x^2\pm\sqrt{\frac{\Delta^2}{4}+(gx-\frac{\epsilon}{2})^2},\\
V_{eff,\pm3}&=&\frac{1}{2}mw_{0}^{2}x^2\pm3\sqrt{\frac{\Delta^2}{4}+(gx-\frac{\epsilon}{2})^2},
\end{eqnarray}
where $V_{eff,\pm1}$ and $V_{eff,\pm3}$ correspond to three qubits
being in the states $|A_{\pm1}\rangle$ and $|A_{\pm3}\rangle$,
respectively. We remark that the effective potentials in Eqs. (20)
and (21) are no longer the harmonic potentials, and the second terms
in Eqs. (20) and (21) represent branches of two kinds of hyperbolas
associated with different states of three qubits, but its analytical
results are mathematically tough to derive generally. Therefore, we
analyze some special cases in the following.

Under the condition $E_{q}>>g|x|$, the effective potentials in Eqs.
(20) and (21) can be approximated by:
\begin{eqnarray}\label{e22-23}
V_{eff,\pm1}&\approx&\frac{1}{2}mw_{0}^{2}x^2\pm\bigg(
\frac{\sqrt{\Delta^2+\epsilon^2}}{2}-\frac{\epsilon
gx-g^2x^2}{\sqrt{\Delta^2+\epsilon^2}}\bigg)\cr\cr
&=&\frac{1}{2}m\tilde{w}^{2}_{0,\pm1}\bigg( x\mp\frac{\epsilon
g}{m\tilde{w}^{2}_{0,\pm1}E_{q}}\bigg)^2\pm\frac{E_{q}}{2},\\ \cr
V_{eff,\pm3}&\approx&\frac{1}{2}
mw_{0}^{2}x^2\pm3\bigg(\frac{\sqrt{\Delta^2+\epsilon^2}}{2}-\frac{\epsilon
gx-g^2x^2}{\sqrt{\Delta^2+\epsilon^2}}\bigg)\cr\cr
&=&\frac{1}{2}m\tilde{w}^{2}_{0,\pm3}\bigg( x\mp\frac{3\epsilon
g}{m\tilde{w}^{2}_{0,\pm3}E_{q}}\bigg)^2\pm\frac{3E_{q}}{2},
\end{eqnarray}
where
\begin{eqnarray}\label{e24-25}
\tilde{w}^{2}_{0,\pm1}&=&w_{0}^{2}\pm2\frac{g^2}{mE_{q}},\\ \cr
\tilde{w}^{2}_{0,\pm3}&=&w_{0}^{2}\pm6\frac{g^2}{mE_{q}}.
\end{eqnarray}
The oscillator's effective potentials have two ways of being
dependent on the states of three qubits. One way is that the
location of the minimum potential deviates from the origin point
with a quantity proportional to $\epsilon/E_{q}$. The other way is
that the oscillator's frequency is renormalized depending on
different states of three qubits: the oscillator's effective
frequency increases when the number of qubits in the excited state
is more than that in the ground state, while the oscillator's
effective frequency is reduced when the number of qubits in the
excited state is less than that in the ground state. Note that the
variation in the renormalized oscillator's frequency caused by the
state of three qubits is proportional to the difference between the
number of qubits in the excited state and that in the ground state,
which is impossible in the system of one oscillator with only one
qubit \cite{PRA-81-042311-2010}.

Especially, when three qubits are in their ground states and
\begin{eqnarray}\label{e26}
\frac{6g^2}{mw_{0}^{2}E_q}>1,
\end{eqnarray}
or one of three qubits is in its excited state and the other two
qubits are in the ground states and
\begin{eqnarray}\label{e27}
\frac{2g^2}{mw_{0}^{2}E_q}>1,
\end{eqnarray}
the renormalized frequency in Eq. (24) or (25) is mathematically
imaginary, indicating that critical points emerge above which the
qubit-oscillator system becomes instable. When
$|x|>>\Delta/g,\epsilon/g$, the effective potentials behave well
with the following ways:
\begin{eqnarray}\label{e28-29}
V_{eff,\pm1}&\simeq&\frac{1}{2}mw_{0}^{2}x^2\pm\big|gx-\frac{\epsilon}{2}\big|,\\
V_{eff,\pm3}&\simeq&\frac{1}{2}mw_{0}^{2}x^2\pm3\big|gx-\frac{\epsilon}{2}\big|.
\end{eqnarray}
Therefore, the assumption of three adiabatically adjusting qubits
above the critical points is still valid.

To see more abundant contents in the ultrastrong-coupling regime,
i.e., above the critical points, we separate the discussions into
two parts: $\epsilon=0$ and $\epsilon\neq0$.

For $\epsilon=0$, the effective potentials in Eqs. (22) and (23)
turn into :
\begin{eqnarray}\label{e30}
 V_{eff,\pm1}=
  \left\{
   \begin{array}{l}
   \big( \frac{1}{2}mw_{0}^{2}\pm\frac{g^2}{\Delta}\big)x^{2}\pm\frac{\Delta}{2},
   |x|<<\Delta/g,\\ \cr
   \frac{1}{2}mw_{0}^{2}x^{2}\pm\big| gx\big|, |x|>>\Delta/g,
   \end{array}
  \right.
\end{eqnarray}
and
\begin{eqnarray}\label{e31}
 V_{eff,\pm3}=
  \left\{
   \begin{array}{l}
   \big( \frac{1}{2}mw_{0}^{2}\pm\frac{3g^2}{\Delta}\big)x^{2}\pm\frac{\Delta}{2},
   |x|<<\Delta/g,\\ \cr
   \frac{1}{2}mw_{0}^{2}x^{2}\pm3\big| gx\big|, |x|>>\Delta/g,
   \end{array}
  \right.
\end{eqnarray}
respectively. When three qubits are in the state $|A_{-3}\rangle$
and with $6g^2>mw_{0}^2\Delta$, or in the state $|A_{-1}\rangle$ and
with $2g^2>mw_{0}^2\Delta$, say, above the critical point,
$V_{eff,-1}$ or $V_{eff,-3}$ becomes a double-well potential. The
locations of the minimum potential can be found through the
mathematical derivations $\mathrm{d} V_{eff,-1}/\mathrm{d}x=0$ and
$\mathrm{d} V_{eff,-3}/\mathrm{d}x=0$ :
\begin{eqnarray}\label{e32-33}
x_{0,-1}&=&\pm\sqrt{\frac{g^2}{m^2w_{0}^4}-\frac{\Delta^2}{4g^2}},\\
x_{0,-3}&=&\pm\sqrt{\frac{9g^2}{m^2w_{0}^4}-\frac{\Delta^2}{4g^2}}.
\end{eqnarray}
Above the corresponding critical point, the locations of the minimum
potential become :
\begin{eqnarray}\label{e34-35}
x_{0,-1}&\simeq&\pm\frac{g}{mw_{0}^2},\\
x_{0,-3}&\simeq&\pm\frac{3g}{mw_{0}^2},
\end{eqnarray}
with the minimum potential energies :
\begin{eqnarray}\label{e36-37}
V_{min,-1}=V_{eff,-1}(
x_{0,-1})-V_{eff,-1}(0)\approx-\frac{g^2}{2mw_{0}^{2}},\cr \\
V_{min,-3}=V_{eff,-3}(
x_{0,-3})-V_{eff,-3}(0)\approx-\frac{9g^2}{2mw_{0}^{2}},\cr
\end{eqnarray}
and the curvatures for both $V_{eff,-1}$ and $V_{eff,-3}$ are
approximately equal to each other :
\begin{eqnarray}\label{e38}
\frac{d^2V_{eff,-1}}{dx^2}\bigg|_{x=\pm
x_{0,-1}}&\simeq&\frac{d^2V_{eff,-3}}{dx^2}\bigg|_{x=\pm
x_{0,-3}}\simeq mw_{0}^2,\cr&&
\end{eqnarray}
which is also equal to that of the free oscillator with $g=0$ and
that of the system with one qubit and one oscillator
\cite{PRA-81-042311-2010}. Based on Eqs. (36) and (37), we can
estimate the energy space between the ground state and the first
excited state above the corresponding critical point.

For $\epsilon\neq0$, if the qubit-oscillator system is far below the
critical points, from Eqs. (22) and (23), we suggest that $\epsilon$
slightly shifts the locations of the minimum in the single-well
potenticals to the left or right; if the qubit-oscillator system is
far above the critical points, $\epsilon$ breaks the symmetry in the
double-well potentials of the slow oscillator, inducing an energetic
tendency to one of the four wells.

\section{Numerical Verification}

In this section, we perform the exactly numerical simulation for the
system with three qubits and an oscillator under special regimes
(i.e., $\hbar w_0>>E_{q}$ and $\hbar w_0<<E_{q}$) to testify their
properties of the energy-level spectrum and the ground state. The
regime with the resonant situation (i.e., $\hbar w_0=E_{q}$) will
also be considered.

\subsection{Energy-level spectrum}

\begin{figure}
\center
  \includegraphics[width=0.9\columnwidth]{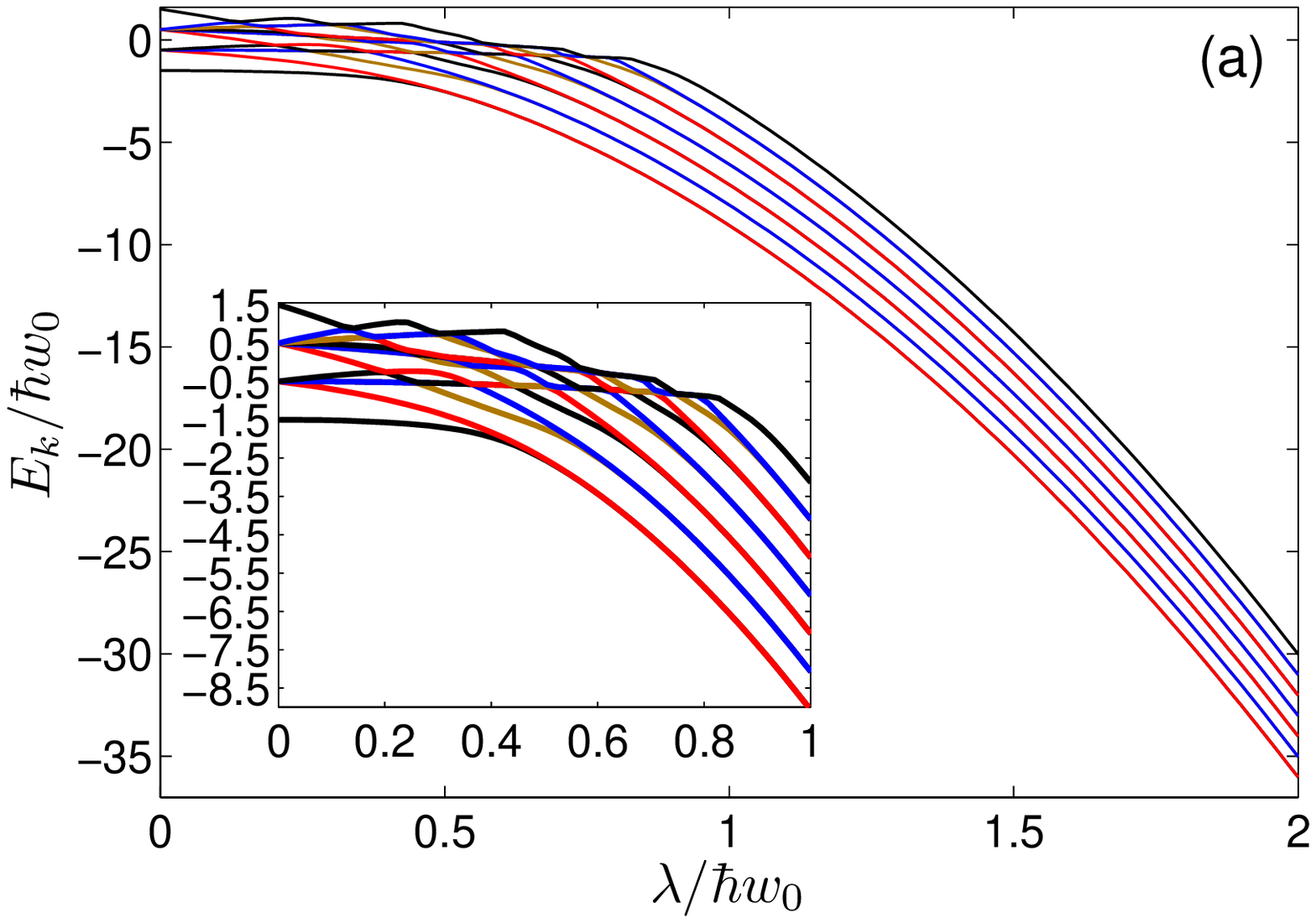}\\
  \includegraphics[width=0.9\columnwidth]{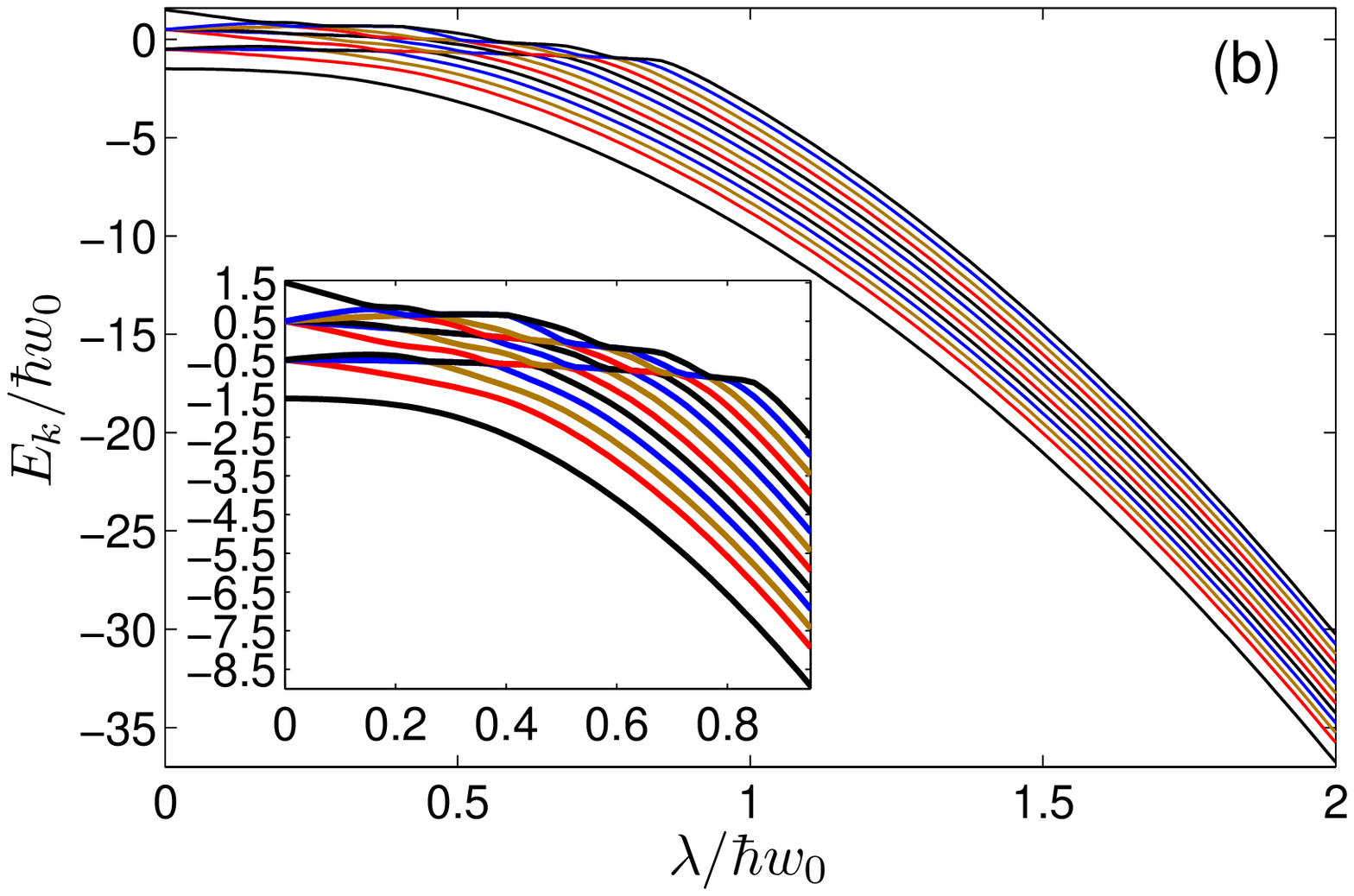}\\
  \includegraphics[width=0.9\columnwidth]{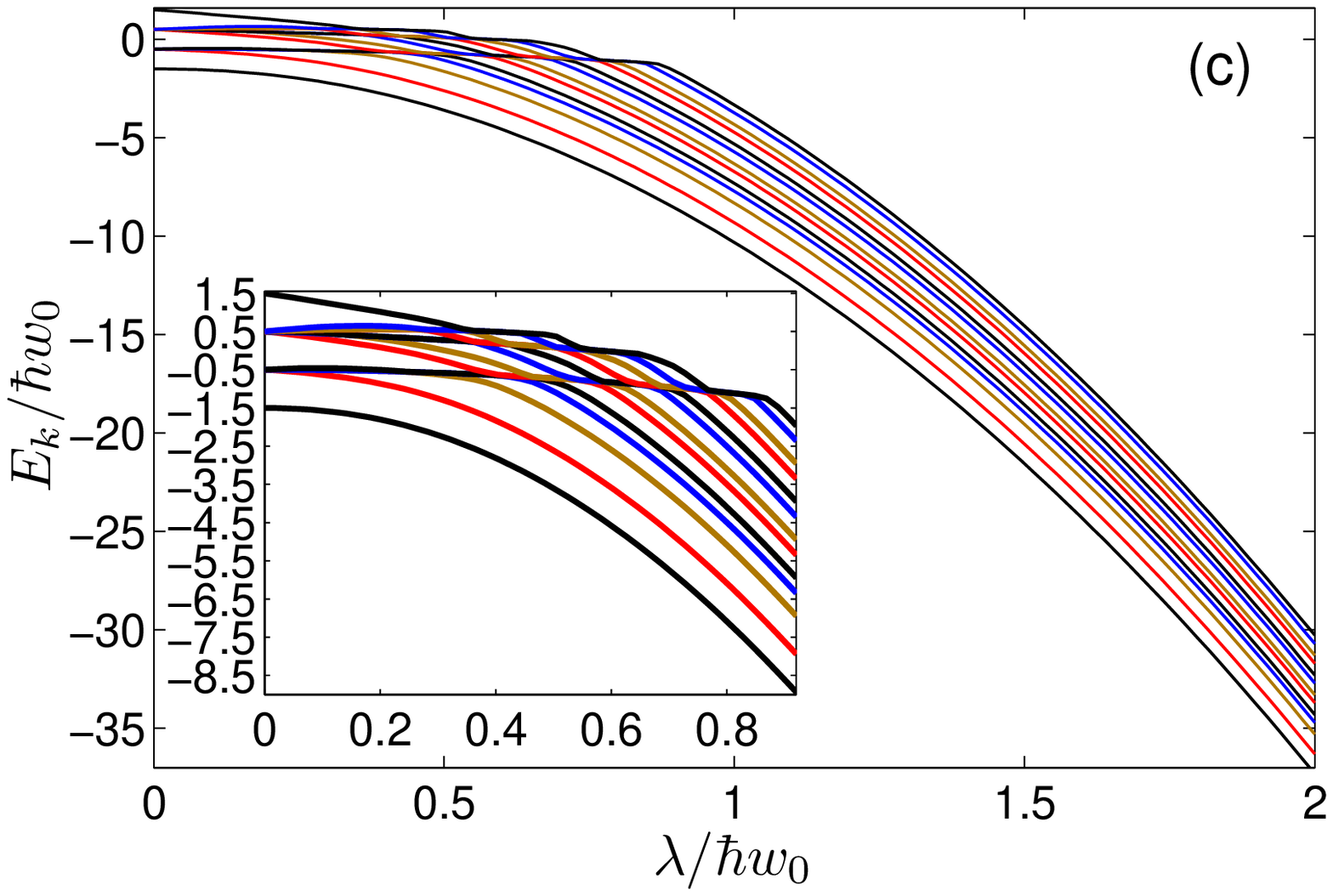} \caption{(Color
  online) Energy spectrums for lowest thirteen levels under the
  resonant situation: $\hbar w_0$/$E_q=1$. The rescaled energy $E_{k}/(\hbar w_0)$ with
  $k=1,2,3,...,13$ versus the rescaled coupling strength $\lambda/(\hbar
  w_0)$ is plotted: (a) $\theta=0$; (b) $\theta=\pi/6$; (c)
  $\theta=\pi/3$.
  }\label{Fig.3.}
\end{figure}

The energies of the lowest thirteen levels versus the
qubit-oscillator coupling strength under the resonant situation
(i.e., $\hbar w_{0}=E_{q}$) are plotted in Fig. 3. In Fig. 3(a), we
find when $\epsilon=\lambda=0$, the ground state is nondegenerate,
and degeneracy degrees of the excited state from being low to high
correspond to 4, 7, 8, 8, ... (the symbol ``...'' represents 8 all
the time as the number of energy level increases which is not
plotted here), respectively, and the space between neighboring
energy levels is $\hbar w_0$. If $\lambda$ increases, the energy
levels move up or down, and all energy levels form doubly degenerate
including the ground state when $\lambda$ becomes large enough, in
which the space between neighboring energy levels is $\hbar w_0$
again. This result coincides with the symmetric structure of
effective double-well potentials in Sec. IIIB. In Fig. 3 (b) and
(c), we see when there is a bias $\epsilon$ (i.e., $\theta\neq0$)
and $\lambda$ is small, the energy levels still keep highly
degenerate which is similar to the case of $\epsilon=0$ in Fig.
3(a). However, when $\lambda$ is large enough, the degeneracy of the
energy levels vanishes and the space between neighboring energy
levels varies with $\epsilon$. This variation in energy level caused
by $\epsilon$ indicates the asymmetry structure in effective
double-well potentials.

\begin{figure}
\center
  \includegraphics[width=0.9\columnwidth]{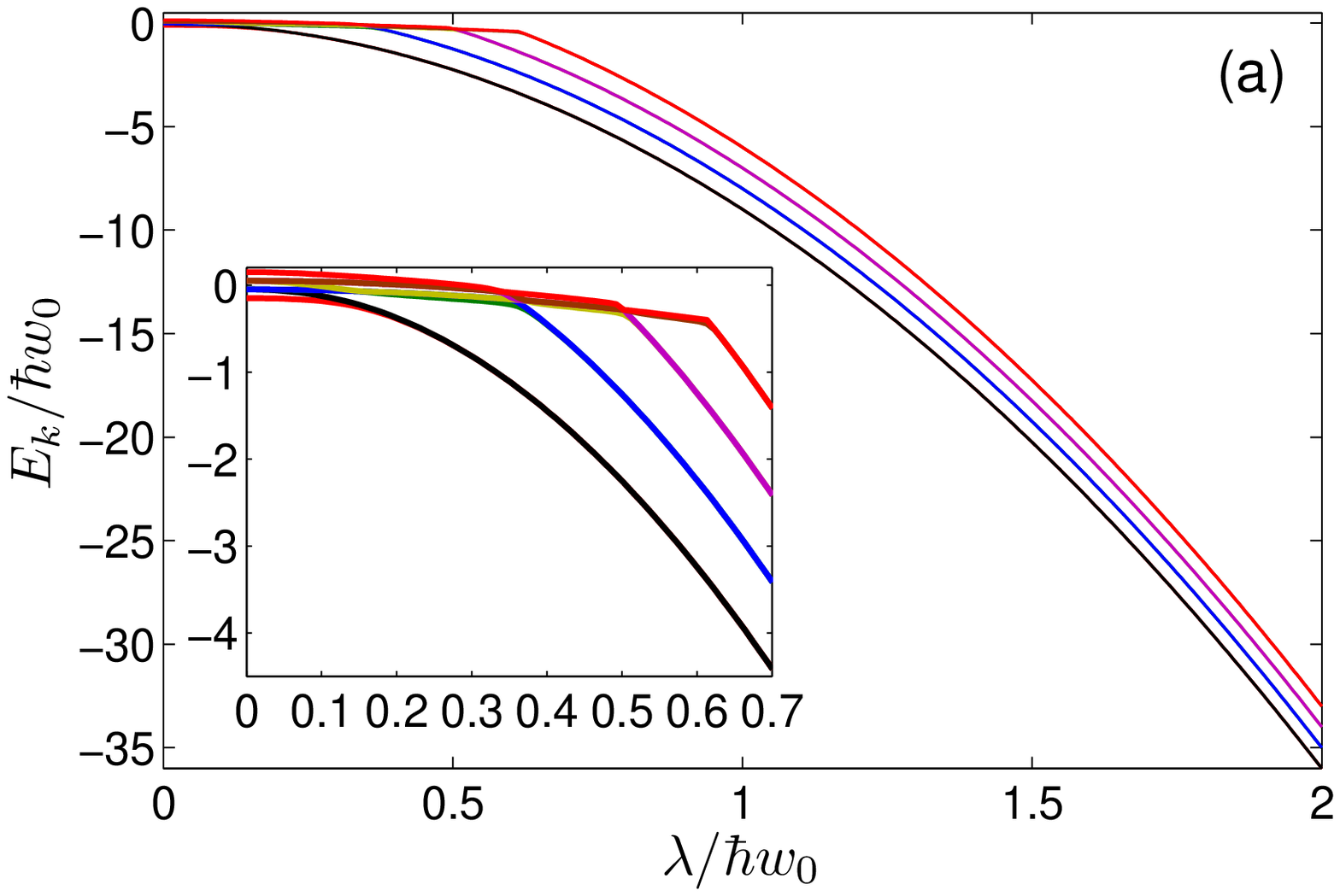}\\
  \includegraphics[width=0.9\columnwidth]{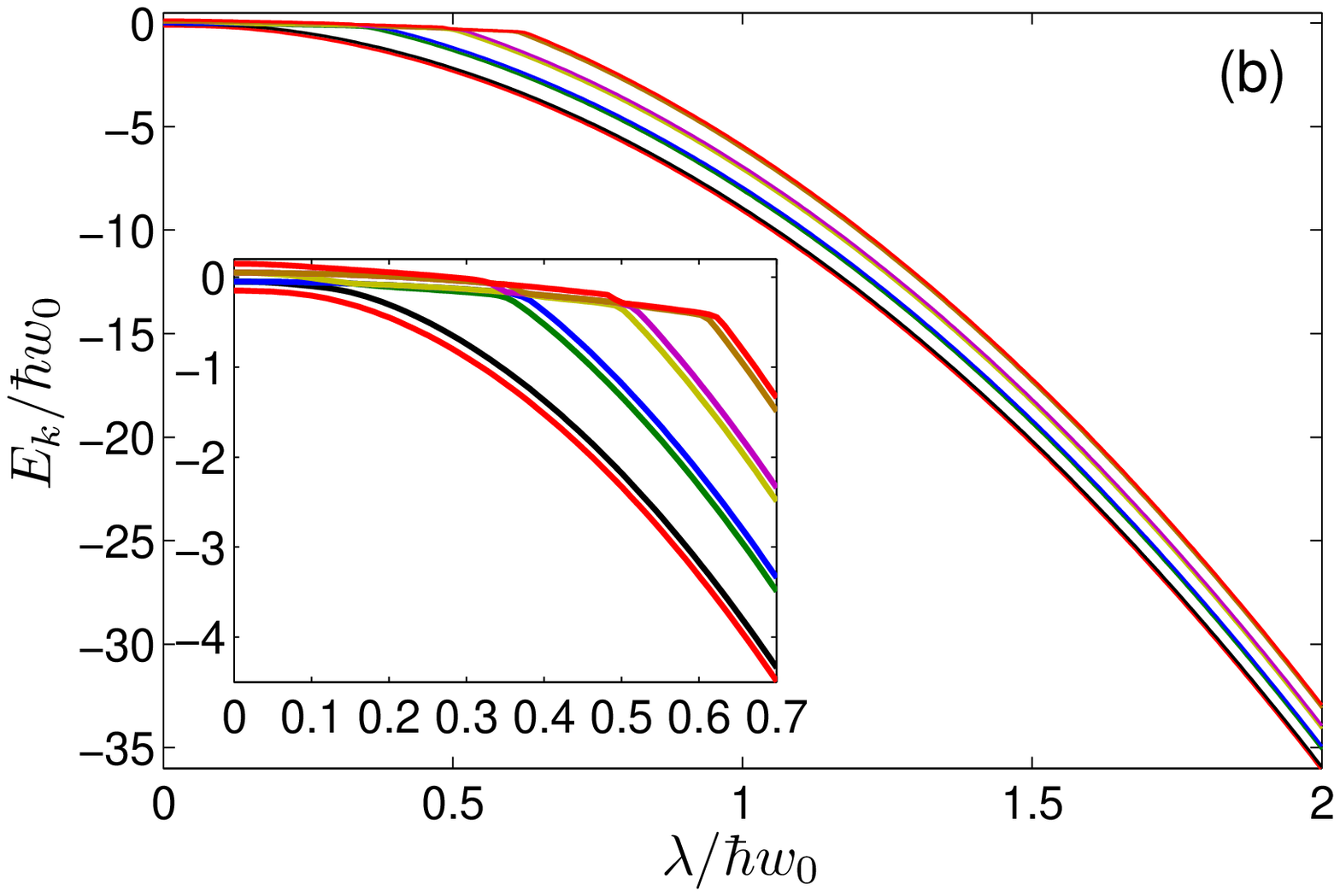}\\
  \includegraphics[width=0.9\columnwidth]{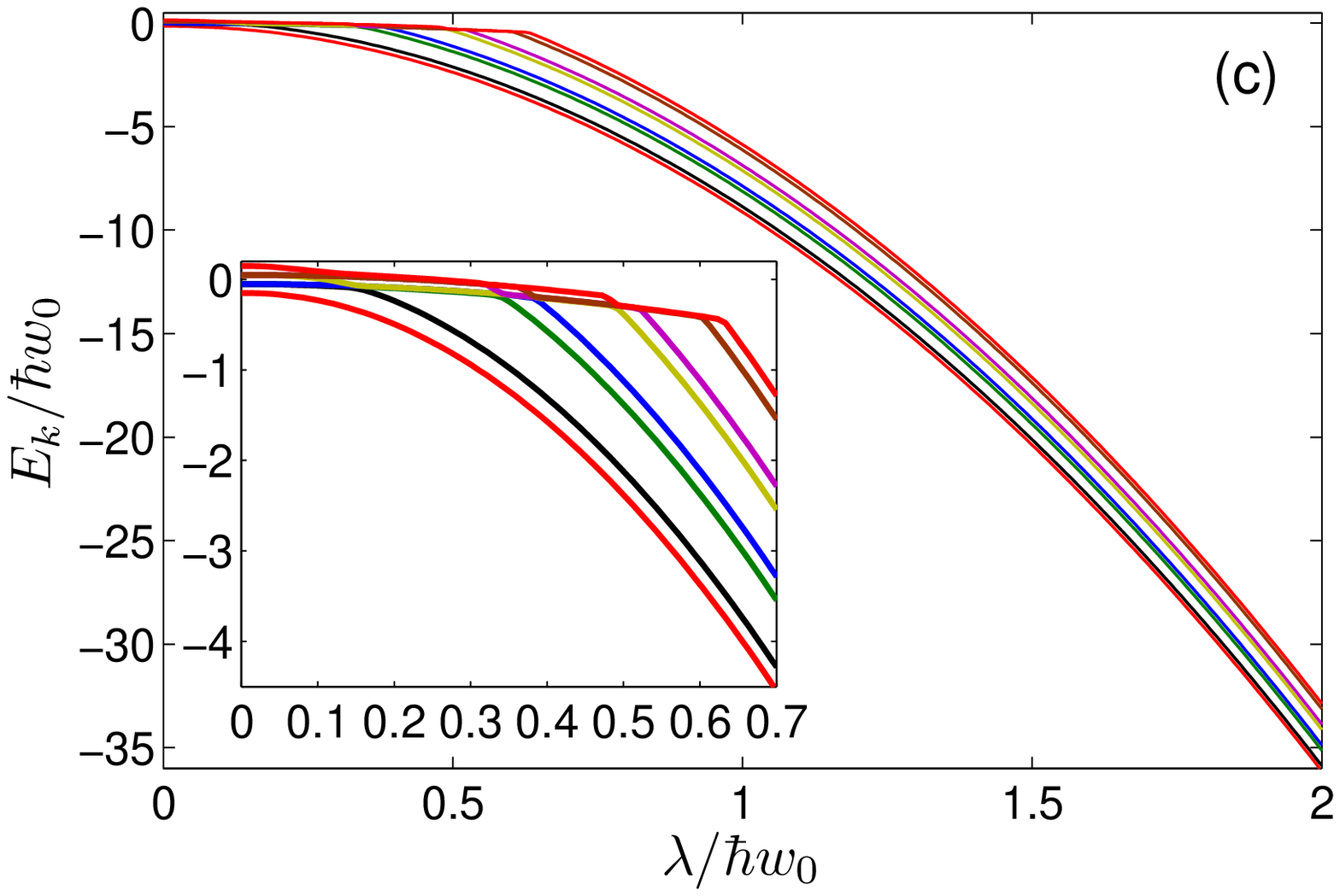} \caption{(Color
  online) Energy spectrums for lowest eight levels under the situation with a
  high-frequency oscillator: $\hbar w_0$/$E_q=10$. The rescaled energy
  $E_{k}/(\hbar w_0)$ with
  $k=1,2,3,...,8$ versus the rescaled coupling strength $\lambda/(\hbar
  w_0)$ is plotted: (a) $\theta=0$; (b) $\theta=\pi/6$; (c)
  $\theta=\pi/3$.
  }\label{Fig.4.}
\end{figure}

In Fig. 4, we plot the spectrums of the lowest eight energy levels
versus the qubit-oscillator coupling strength under the situation
with a high-frequency oscillator (i.e., $\hbar w_0>>E_q$). We find
that, when $\epsilon=0$, all energy levels form pairs and become
doubly degenerate as $\lambda$ becomes large enough, of which the
energy gap is $\hbar w_0$; when $\epsilon\neq0$ and $\lambda$ is
large enough, the degeneracy in energy-level pairs vanishes and
neighboring energy levels are separated by a quantity approximate to
$3\epsilon$. This coincides with the result derived in Sec. IIIA.
For each value of $n$ and the large-$\lambda$ limit, the
oscillator's effective eigenenergy in Eq. (14) becomes a dominant
component in the low energy levels, i.e., three qubits are in the
states $|A_{\pm3}\rangle$ for the large-$\lambda$ limit, producing
the effective energy gap $\Delta_{eff}$ $=$
$3\Delta\sqrt{l^2+\tan^2\theta}\simeq3\epsilon$ ($l\rightarrow0$ for
large-$\lambda$ limit) in energy-level pairs. This is very different
from that in the system of one oscillator with only one qubit
\cite{PRA-81-042311-2010}, in which the spaces between low energy
levels are independent of the bias $\epsilon$.

\begin{figure}
\center
  \includegraphics[width=0.9\columnwidth]{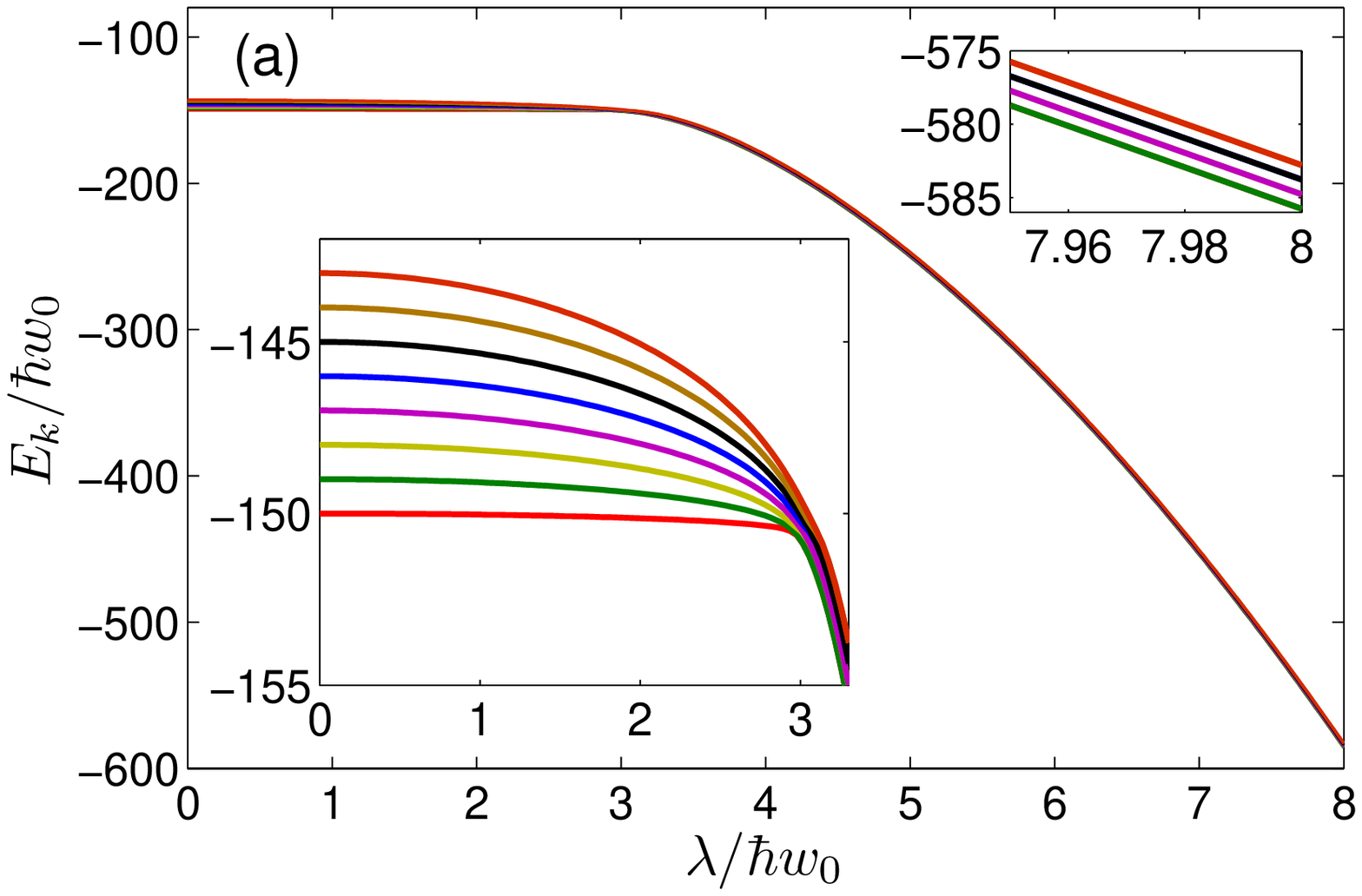}\\
  \includegraphics[width=0.9\columnwidth]{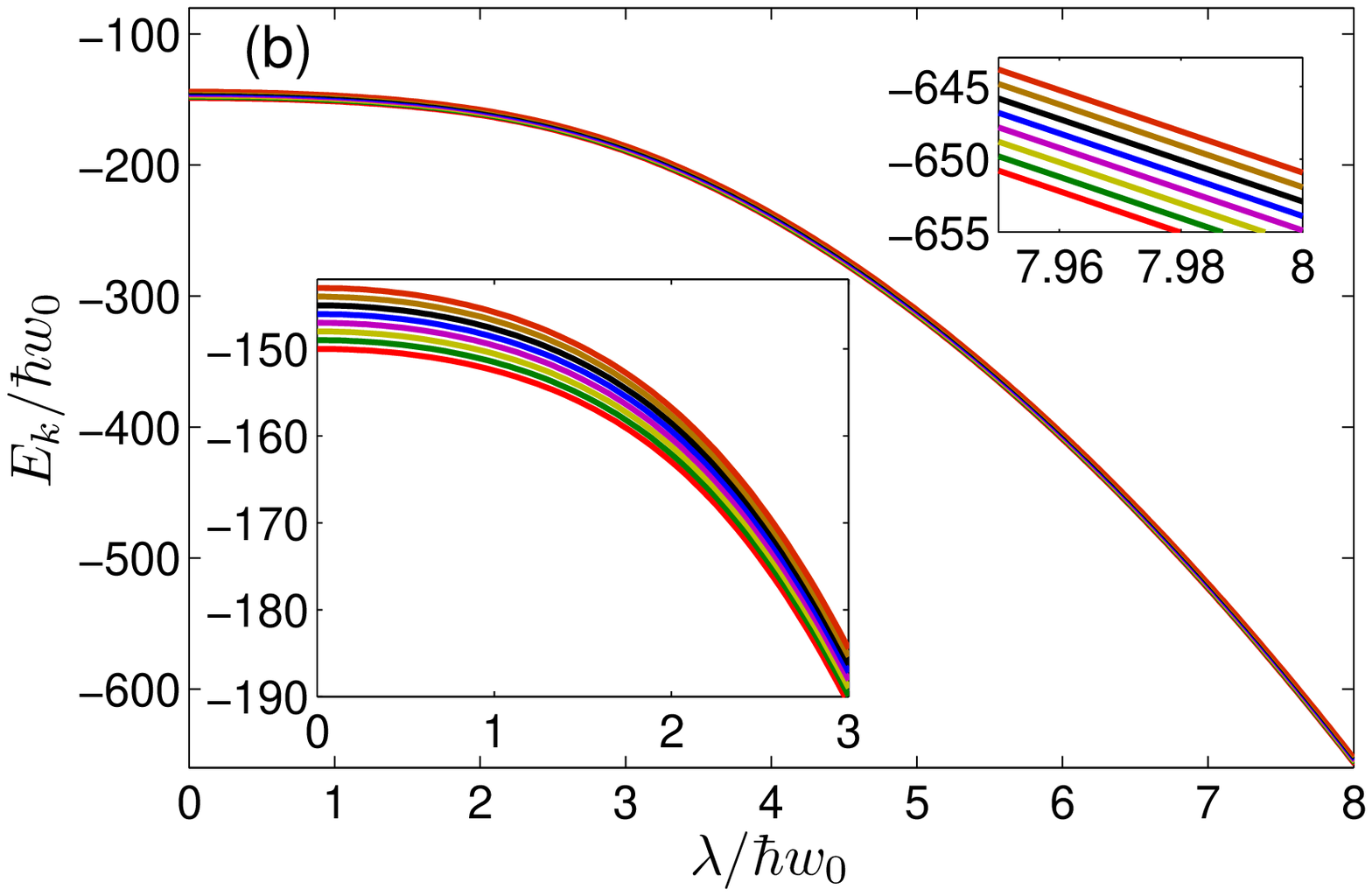}\\
  \includegraphics[width=0.9\columnwidth]{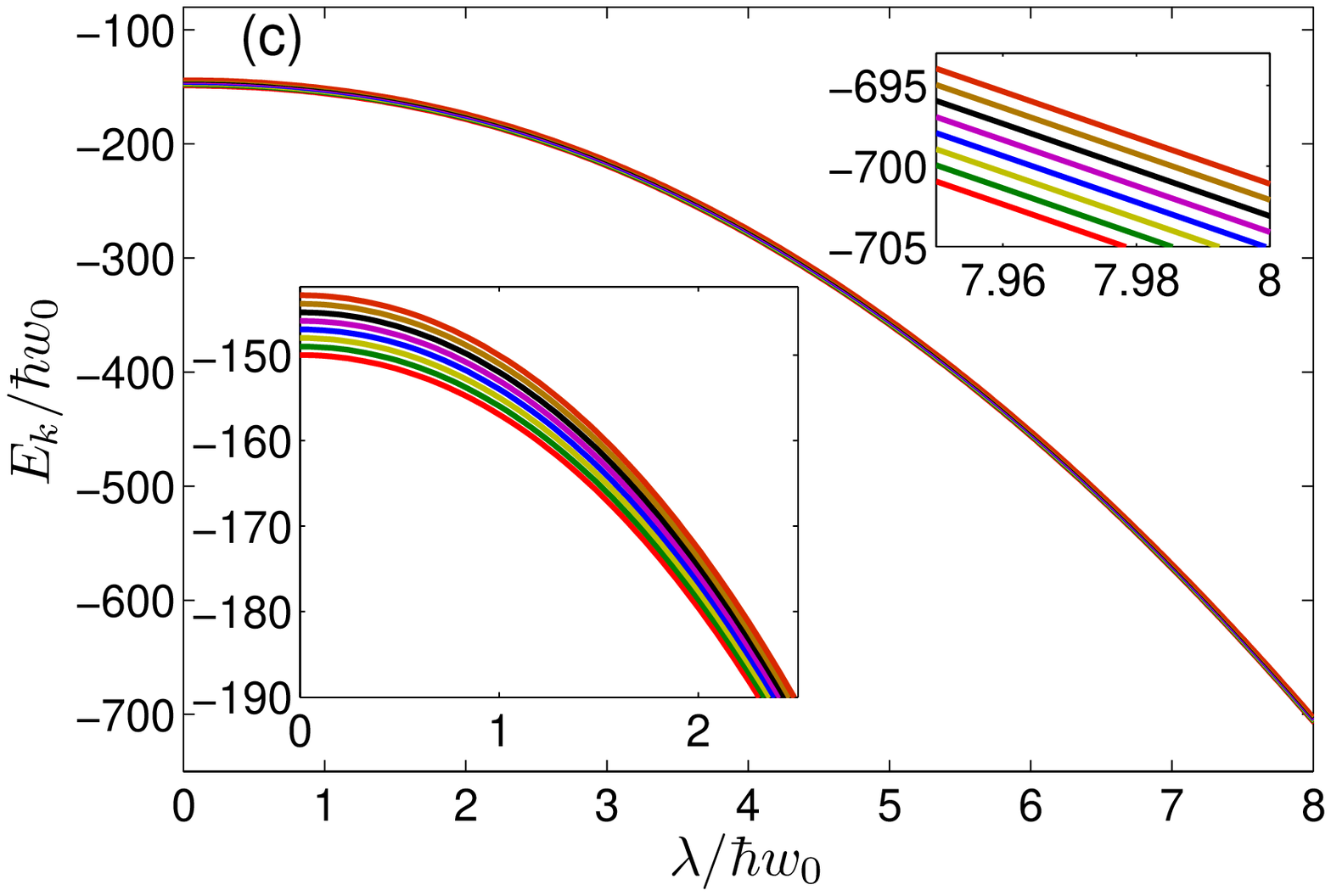} \caption{(Color
  online) Energy spectrums for lowest eight levels under the situation with three
  high-frequency qubits: $\hbar w_0$/$E_q=0.01$. The rescaled
  energy $E_{k}/(\hbar w_0)$ with
  $k=1,2,3,...,8$ versus the rescaled coupling strength $\lambda/(\hbar
  w_0)$ is plotted: (a) $\theta=0$; (b) $\theta=\pi/6$; (c)
  $\theta=\pi/3$.
  }\label{Fig.5.}
\end{figure}

In Fig. 5, we plot the spectrums of the lowest eight energy levels
versus the qubit-oscillator coupling strength under the situation
with three high-frequency qubits (i.e., $E_{q}>>\hbar w_0$). The
most interesting effect emerges in the case with $\theta=0$. The
ground-state energy maintains constant for
$0\leq\lambda\leq\lambda_c$, where $\lambda_c=\sqrt{\hbar
w_0\Delta}/(2\sqrt{N_j})$ ($N_j=3$) is the critical point predicted
for a vanishing $\tilde{w}_{0,\pm3}$. When $\lambda$ $>$
$\lambda_c$, the ground-state energy decreases indefinitely with
increasing $\lambda$. Especially, the lowest eight energy levels
come close to each other with increasing $\lambda$ below the
critical point $\lambda_c$, and become doubly degenerate when
$\lambda$ is beyond $\lambda_c$. For the case with $\theta\neq0$,
there is no critical point in the ground-state energy any longer,
and the lowest eight energy levels do not form doubly degenerate
with increasing $\lambda$, between which the spaces are independent
of $\theta$.

\subsection{Properties of ground state}

In this section, we analyze the nonclassical properties of the
ground state in our qubit-oscillator system under various
combinations of system parameters, which are necessary for preparing
the oscillator squeezed state, the oscillator Schr\"{o}dinger-cat
state, the qubit-qubit entangled state, and the qubit-oscillator
entangled state in the ultrastrong coupling regime.

\begin{figure}
\center
  \includegraphics[width=1\columnwidth]{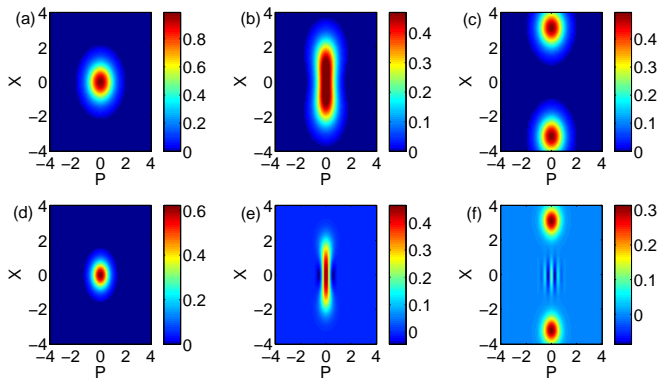} \caption{(Color
  online) The $Q$ function (upside) and the Wigner function
  (underside) of the oscillator's state with three high-frequency
  qubits (i.e., $\hbar w_0/\Delta=0.1$ and $\epsilon=0$): (a,d) $\lambda/(\hbar
  w_0)=0.5$, (b,e) $\lambda/(\hbar w_0)=1$, (c,f) $\lambda/(\hbar
  w_0)=1.25$.
  }\label{Fig.6.}
\end{figure}

We first consider the $Q$ function and Wigner function of the
oscillator in the ground state of the qubit-oscillator system. The
$Q$ function is defined as :
\begin{eqnarray}\label{e39}
Q(X,P)=\frac{1}{\pi}\langle X-iP|\rho_{osc}|X+iP\rangle,
\end{eqnarray}
where $\rho_{osc}=Tr_q\{|\Psi_{GS}\rangle\langle\Psi_{GS}|\}$ is the
oscillator's reduced density matrix through tracing out three qubits
from the ground state $|\Psi_{GS}\rangle$, and $|X\pm iP\rangle$
represent the coherent states $|\alpha\rangle$ with complex
amplitudes $\alpha=X\pm iP$. The Wigner function is defined as :
\begin{eqnarray}\label{e40}
W(X,P)&=&\frac{1}{2\pi}\int_{-\infty}^{\infty}\langle
X+\frac{1}{2}X^{\prime}|\rho_{osc}|X-\frac{1}{2}X^{\prime}\rangle
e^{iPX^{\prime}}dX^{\prime},\cr&&
\end{eqnarray}
where $|X\pm\frac{1}{2}X^{\prime}\rangle$ represents the eigenstates
of the position operator. We plot the $Q$ and Wigner functions
versus parameters $X$ and $P$ under different system parameters in
Fig. 6. The results illustrate that the oscillator's state is
initially in a coherent state with no photons (the vacuum state),
and becomes a qubit-oscillator entangled state with increasing
$\lambda$. The negative value in the Wigner function implies that
the nonclassical state of the oscillator, and the nonclassical
property of the oscillator becomes more obvious when $\lambda$
increases further, as the obvious interference-fringe-like pattern
plotted in Fig. 6(f). %%%%

The Schr\"{o}dinger-cat-like state only appears when two peaks of
the Wigner function in Fig. 6 separate completely between which
there exist oscillations with alternating positive and negative
values, showing the feature of quantum interference. From Fig. 6(e),
the nonclassical state of the oscillator first appears when
$\lambda\simeq\hbar w_0$ in the three-qubit case, which is about one
half of that in the single-qubit case \cite{PRA-81-042311-2010}.
From Fig. 6(f), the minimum qubit-oscillator coupling strength
needed to produce the Schr\"{o}dinger-cat state for our three-qubit
Dicke model is $\lambda\simeq1.25\hbar w_0$. However, in the
single-qubit case, the minimum qubit-oscillator coupling strength
$\lambda$ needed to produce the Schr\"{o}dinger-cat stat is about
$2.5\hbar w_0$ \cite{PRA-81-042311-2010}, which is about twice as
big as that in our three-qubit case. Different from the system of
one oscillator with only one qubit \cite{PRA-81-042311-2010}, the
qubit-oscillator coupling strength needed here for generating the
nonclassical states of the Schr\"{o}dinger-cat type in the
oscillator is much smaller, indicating that the qubit-oscillator
coupling interaction is collectively enhanced as the number of
qubits increases.

%We also calculate the results for
%other situations with $\epsilon$ (not shown here), and the main
%results are as follows:

\begin{figure}
\center
  \includegraphics[width=0.9\columnwidth]{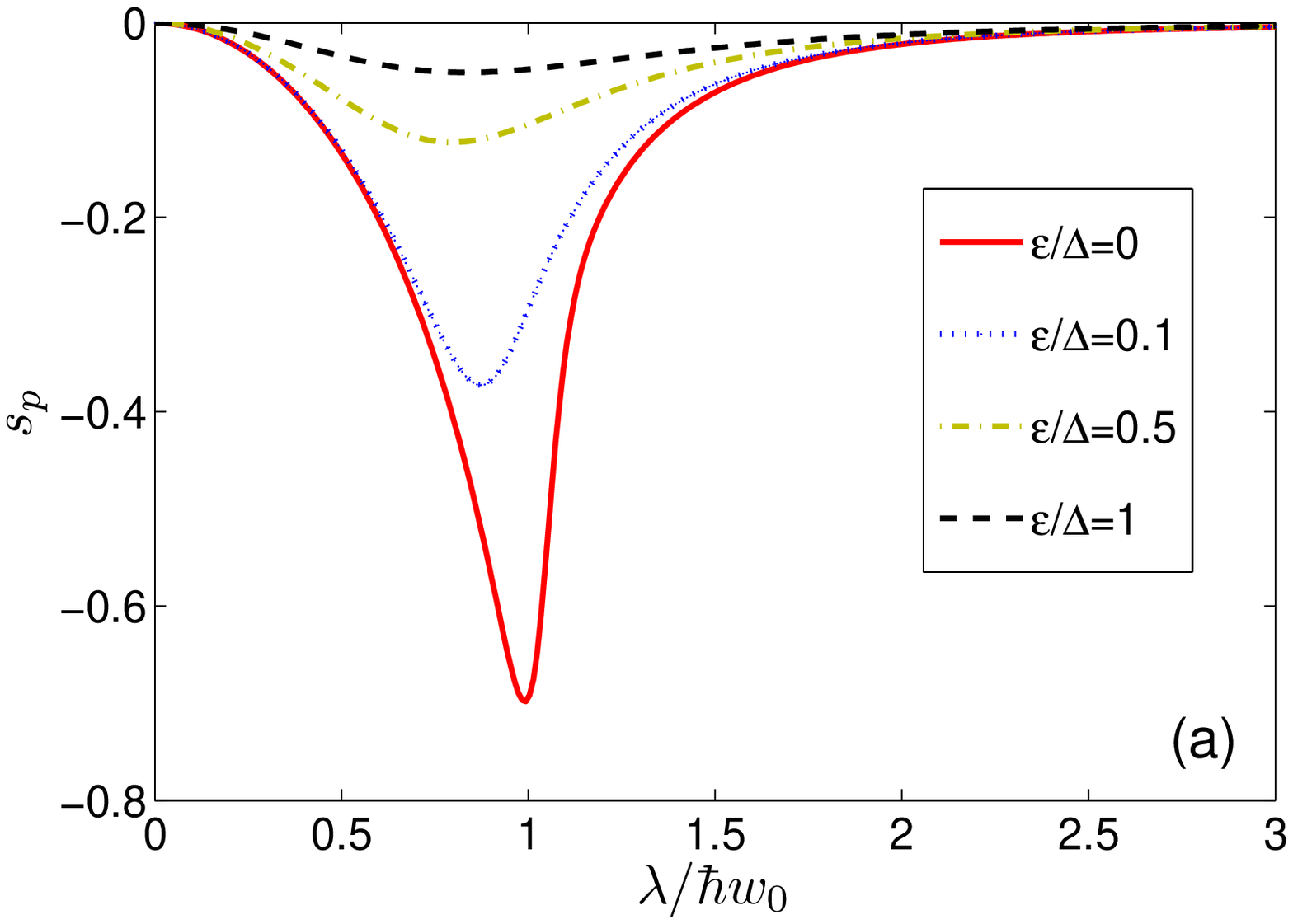}\\
  \includegraphics[width=0.9\columnwidth]{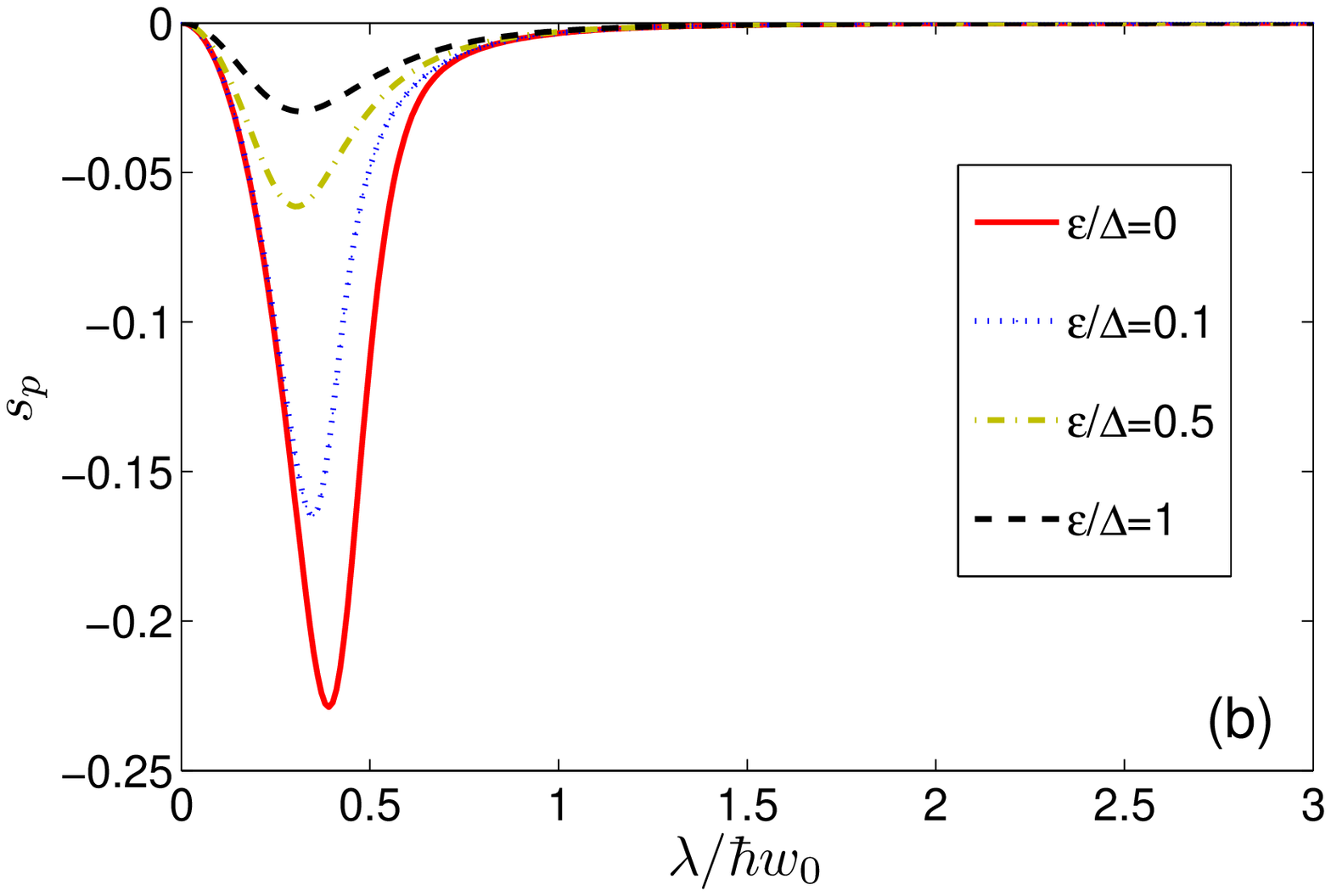}\\
  \includegraphics[width=0.9\columnwidth]{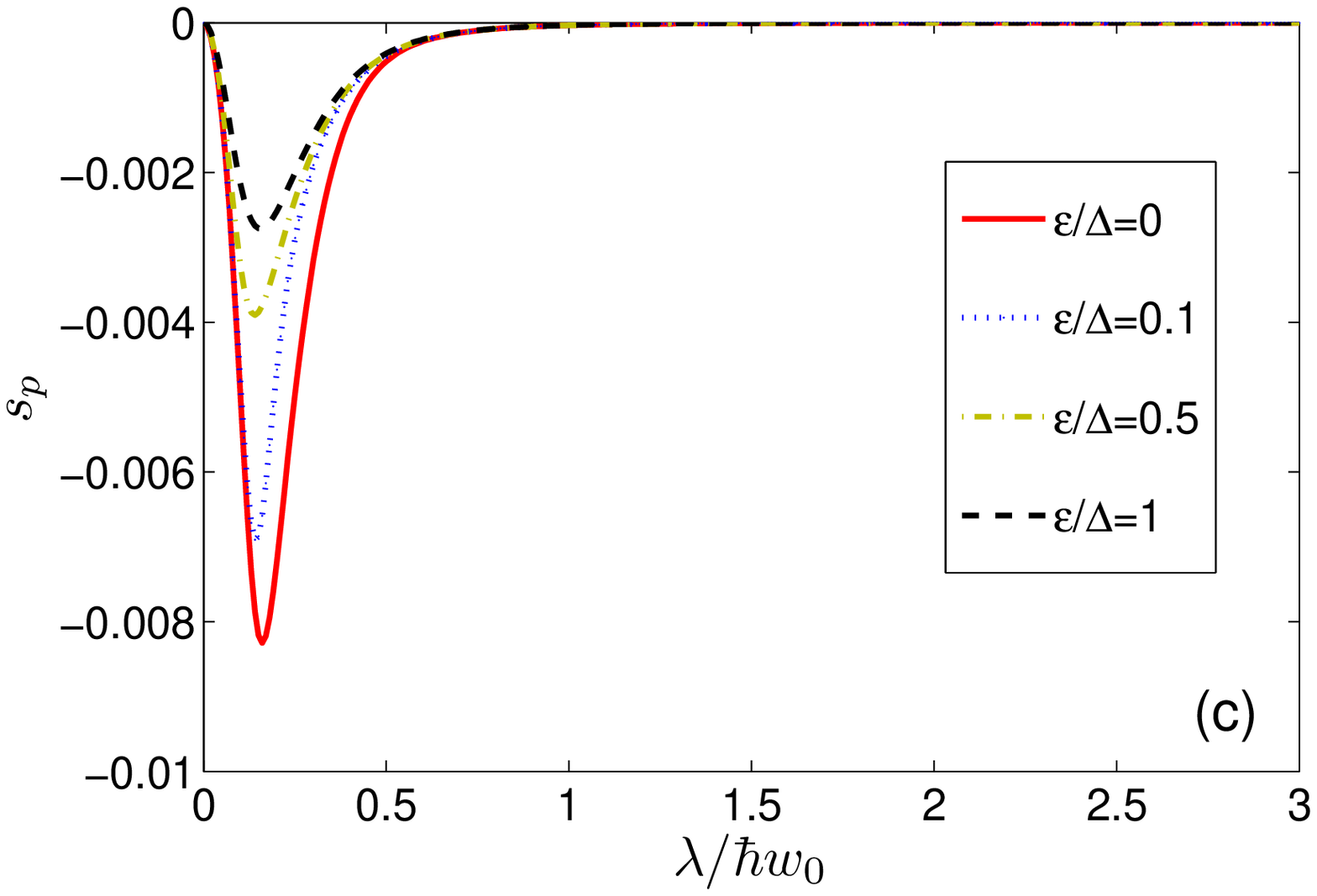} \caption{(Color
  online) The momentum-squeezing
parameter $s_p$ versus $\lambda/(\hbar w_0)$: (a) $\hbar
w_0/\Delta=0.1$, (b) $\hbar w_0/\Delta=1$, (c) $\hbar
w_0/\Delta=10$. }\label{Fig.7.}
\end{figure}

The intuitive quantifier for the squeezing in the oscillator's state
is the set of two squeezing parameters in the $\hat{x}$ and
$\hat{p}$ quadratures, given by \cite{PRA-81-042311-2010}:
\begin{eqnarray}\label{e41-43}
s_p&=&4\langle(\hat{P}-\langle\hat{P}\rangle)^2\rangle-1,\\ \cr
s_x&=&4\langle(\hat{X}-\langle\hat{X}\rangle)^2\rangle-1,\\ \cr
K&=&\frac{\hbar^2}{4}(1+s_p)(1+s_x),
\end{eqnarray}
where $K$ equals to $\hbar^2/4$ for a minimum-uncertainty state
including the coherent and quadrature-squeezed states, while $K$ is
larger than $\hbar^2/4$ for any other states including the
Schr\"{o}dinger-cat and qubit-oscillator entangled states. The
momentum-squeezing parameter versus the qubit-oscillator coupling
strength under different system parameters is plotted in Fig. 7. The
squeezing in oscillator's state increases with increasing $\lambda$
for small $\lambda$ and reaches a maximum, but returns to zero as
$\lambda$ increases further. This is because the ground state gets
entangled in the ultrastrong coupling regime vanishing the
squeezing. The maximum attainable squeezing is the biggest under the
extreme mechanism $\Delta>>\hbar w_0$ and decreases with increasing
ratios $\hbar w_0/\Delta$ and $\epsilon/\Delta$. We also obtain the
numerical results for the parameter $K$, the behaviors of which
versus the qubit-oscillator coupling strength are similar to that
described in the system of one oscillator with only one qubit
\cite{PRA-81-042311-2010}.

Based on the above results, we know that the squeezed state can be
achieved for moderate qubit-oscillator coupling, and the
nonclassical state of the Schr\"{o}dinger-cat type appears when the
Wigner function is negative for relatively strong qubit-oscillator
coupling. To make a clear distinction between the
Schr\"{o}dinger-cat state of the oscillator and the qubit-oscillator
entangled state, we then take further measures to analyze the
properties of the ground-state entanglement.

\begin{figure}
\center
  \includegraphics[width=0.8\columnwidth]{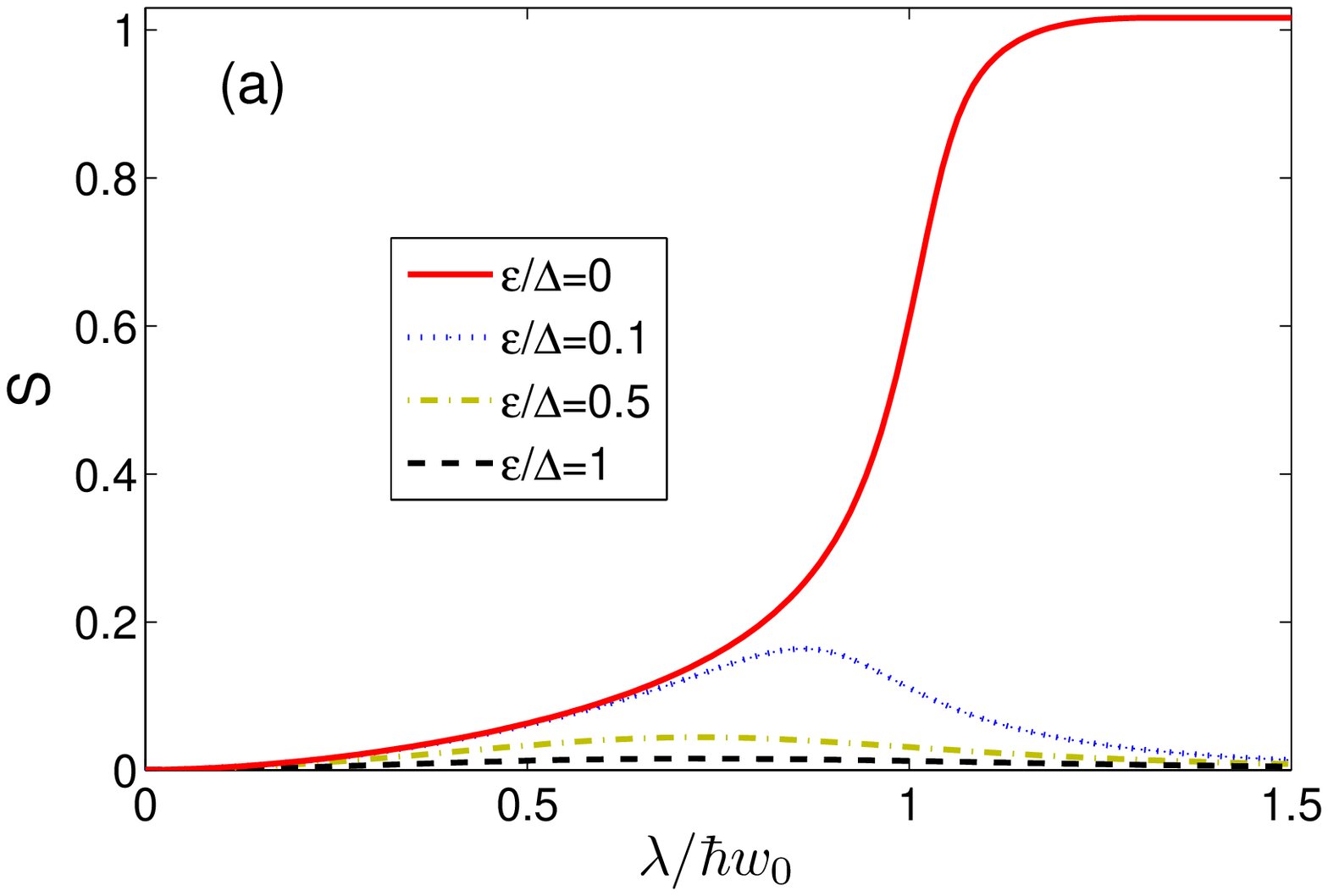}\\
  \includegraphics[width=0.8\columnwidth]{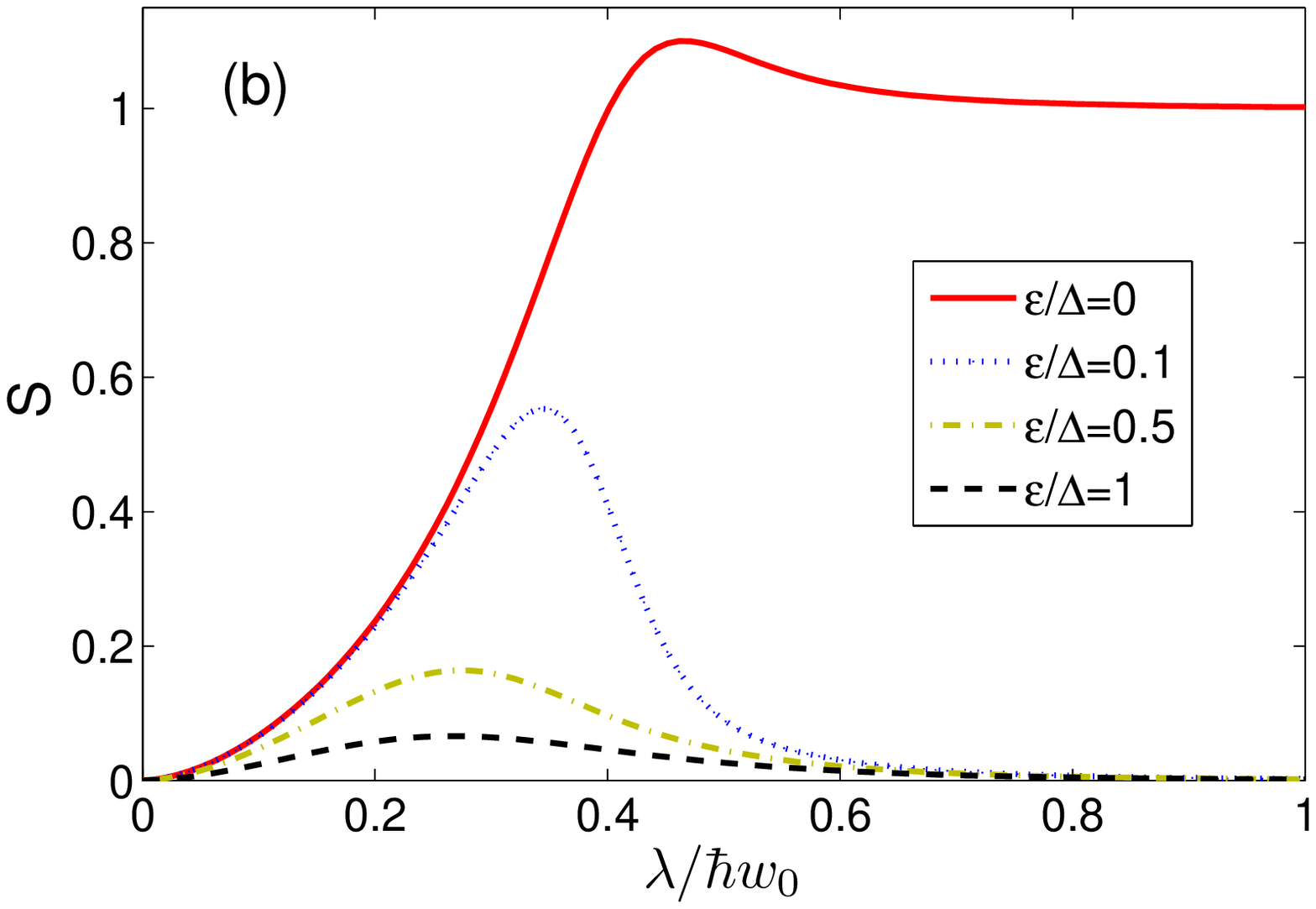}\\
  \includegraphics[width=0.8\columnwidth]{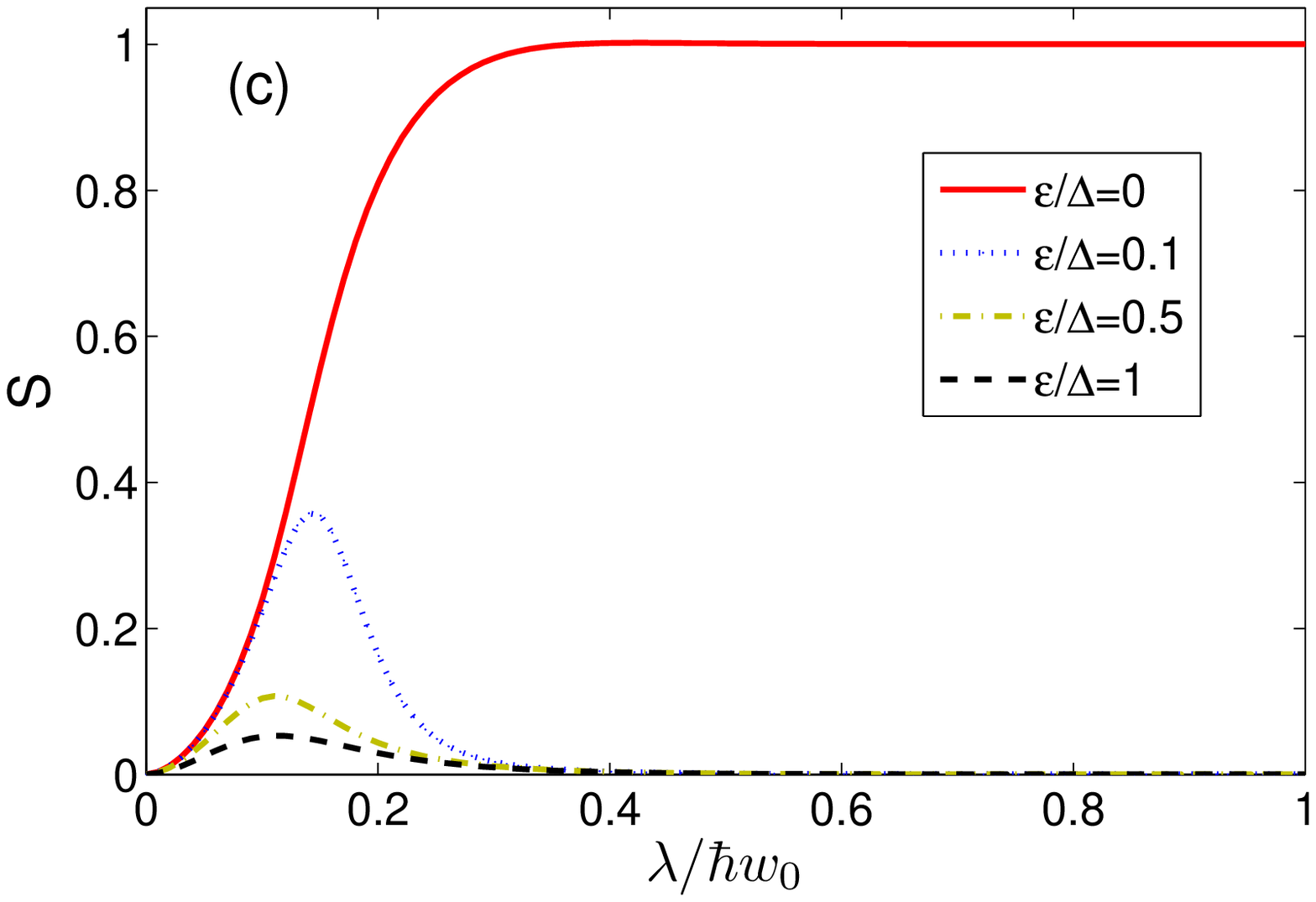}\\
  \includegraphics[width=0.8\columnwidth]{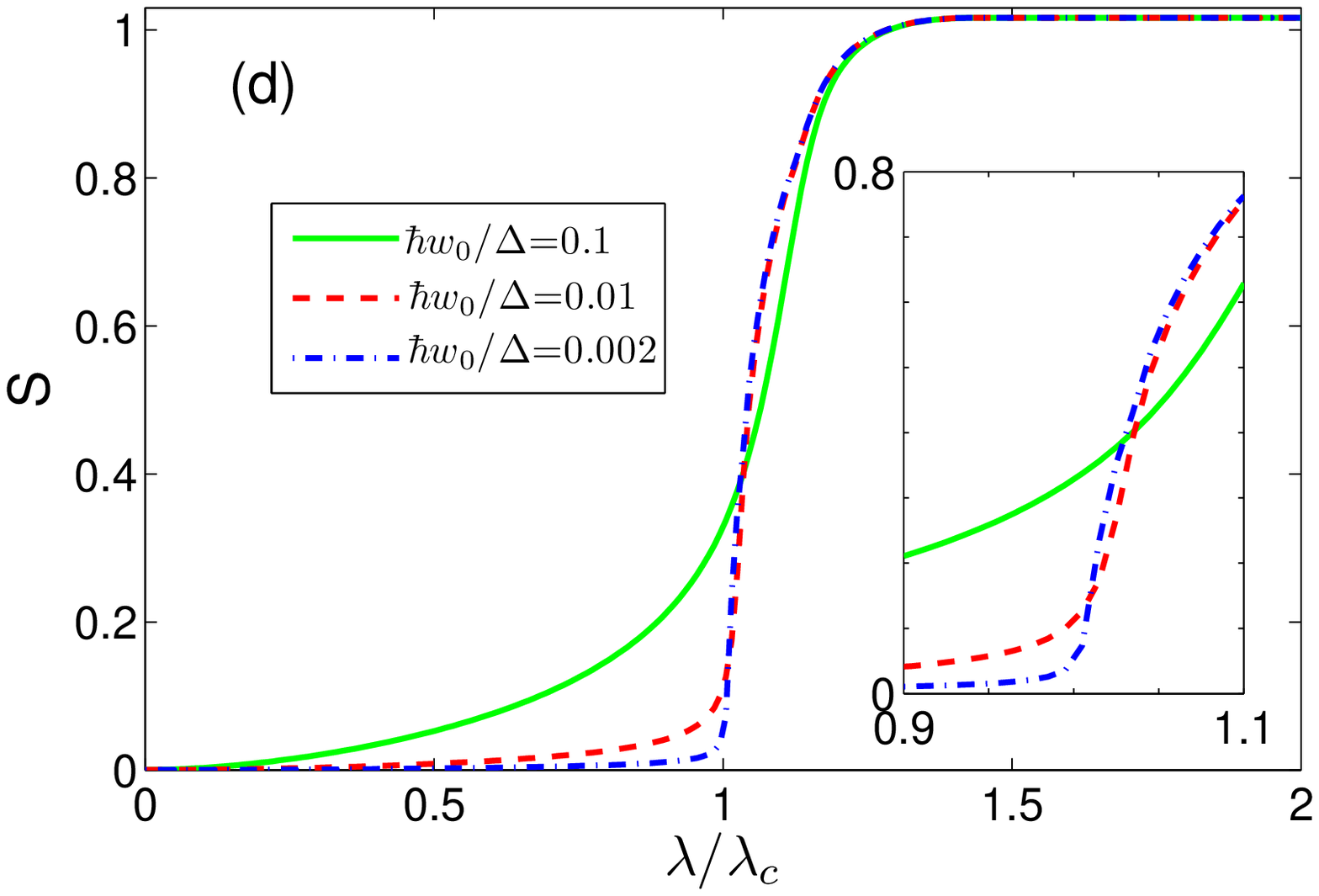} \caption{(Color
  online) The qubits' entropy $S$ versus $\lambda/(\hbar w_0)$ with:
  (a) $\hbar w_0/\Delta=0.1$, (b) $\hbar
  w_0/\Delta=1$, (c) $\hbar w_0/\Delta=10$. (d)
  $S$ versus $\lambda/\lambda_c$, where $\lambda_c$ $=$ $\sqrt{\hbar
w_0\Delta}/(2\sqrt{3})$ and $\epsilon=0$.
  }\label{Fig.8.}
\end{figure}

The ground-state entanglement can be quantified by using the von
Neumann entropy $S$ of three qubits, which is calculated via the
formula $S=-Tr\{\rho_{q}log_{2}\rho_q\}$, where $\rho_{q}=Tr_{osc}\{
|\Psi_{GS}\rangle\langle\Psi_{GS}|\}$ represents three qubits'
reduced density matrix through tracing out the oscillator degrees of
freedom. The entropy of three qubits versus the qubit-oscillator
coupling strength is plotted in Fig. 8, which shows when
$\epsilon=0$, the entropy increases from zero to a value larger than
$1$, and the maximum attainable entropy reaches $1.1$ in Fig.
8(b).%%%%%%%%%%%%%

The von Neumann entropy $S$ has a maximum value of $log_{2}D$ in a
$D$-dimensional Hilbert space. In the case of three qubits, i.e., an
eight-dimensional Hilbert space, the maximum value of qubit-entropy
is $S_{max}=3$, which explains the behavior of qubit-entropy being
not bounded by $1$ in Fig. 8 meaning three qubits in the ground
state can become a nonmaximally entangled state in the limit for
large qubit-oscillator coupling. The interaction among these three
qubits intermediated by the photons of the oscillator becomes much
stronger as the qubit-oscillator coupling strength increases,
explaining the increasing behavior of the three-qubit entropy in the
ground state without $\epsilon$. This is different from the system
of one oscillator with only one qubit \cite{PRA-81-042311-2010}, in
which the maximum attainable entropy is $1$.

For $\epsilon\neq0$, the maximum attainable entropy quickly drops to
zero as $\lambda$ becomes large enough, and never increases again
even when $\lambda$ increases further, indicating the fragility of
the ground-state entanglement at such a condition. For the extreme
situation with three high-frequency qubits plotted in Fig. 8(d), we
assume $\lambda_c$ as a reference point for measuring the coupling
strength. Interestingly we find that, the onset of the qubits'
entropy becomes suddenly increasing when $\lambda$ is varied across
the critical point $\lambda/\lambda_c=1$, and the
phase-transition-like \cite{PRA-81-042311-2010-2} curve in the
entropy appears and becomes more sharp as the ratio $\hbar
w_0/\Delta$ is more close to zero, indicating the system is
experiencing a sudden transition from an uncorrelated state to an
intensively correlated one as the qubit-oscillator coupling strength
increases across the critical point. This phase-transition-like
behavior probed by the entropy with only three high-frequency qubits
loosens the requirement of employing an atomic ensemble for studying
the quantum phase transition
\cite{PR-93-99-1954,PRE-67-066203-2003,PRL-90-044101-2003,PRL-92-073602-2004}.
Note that for non-zero $\epsilon$,  one of the wells is ``deeper''
in Fig. 2, it indeed breaks the parity symmetry of our model and it
is this parity that is spontaneously broken in the superradiant
phase of the Dicke model.
\begin{figure}
\center
  \includegraphics[width=0.9\columnwidth]{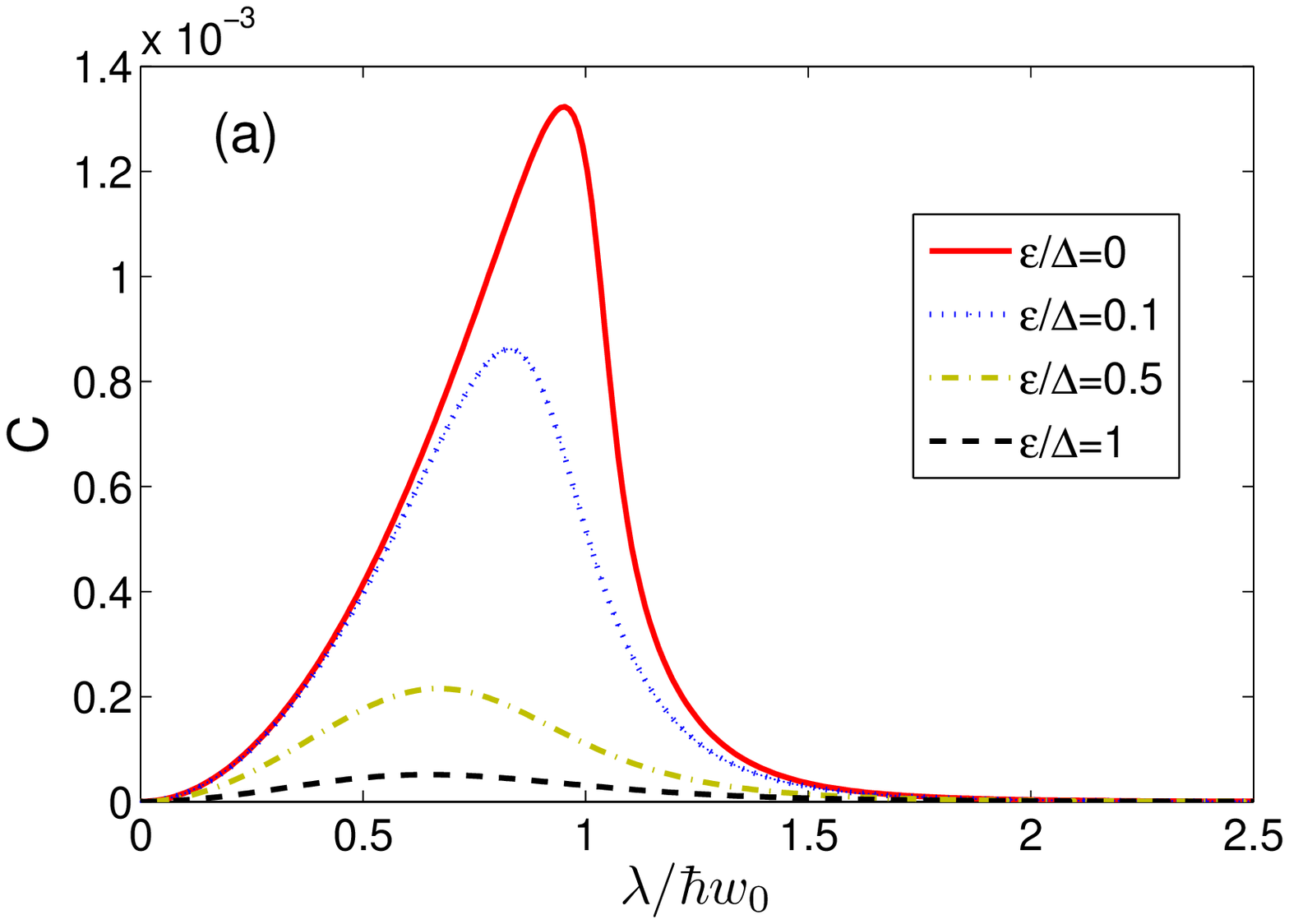}\\
  \includegraphics[width=0.9\columnwidth]{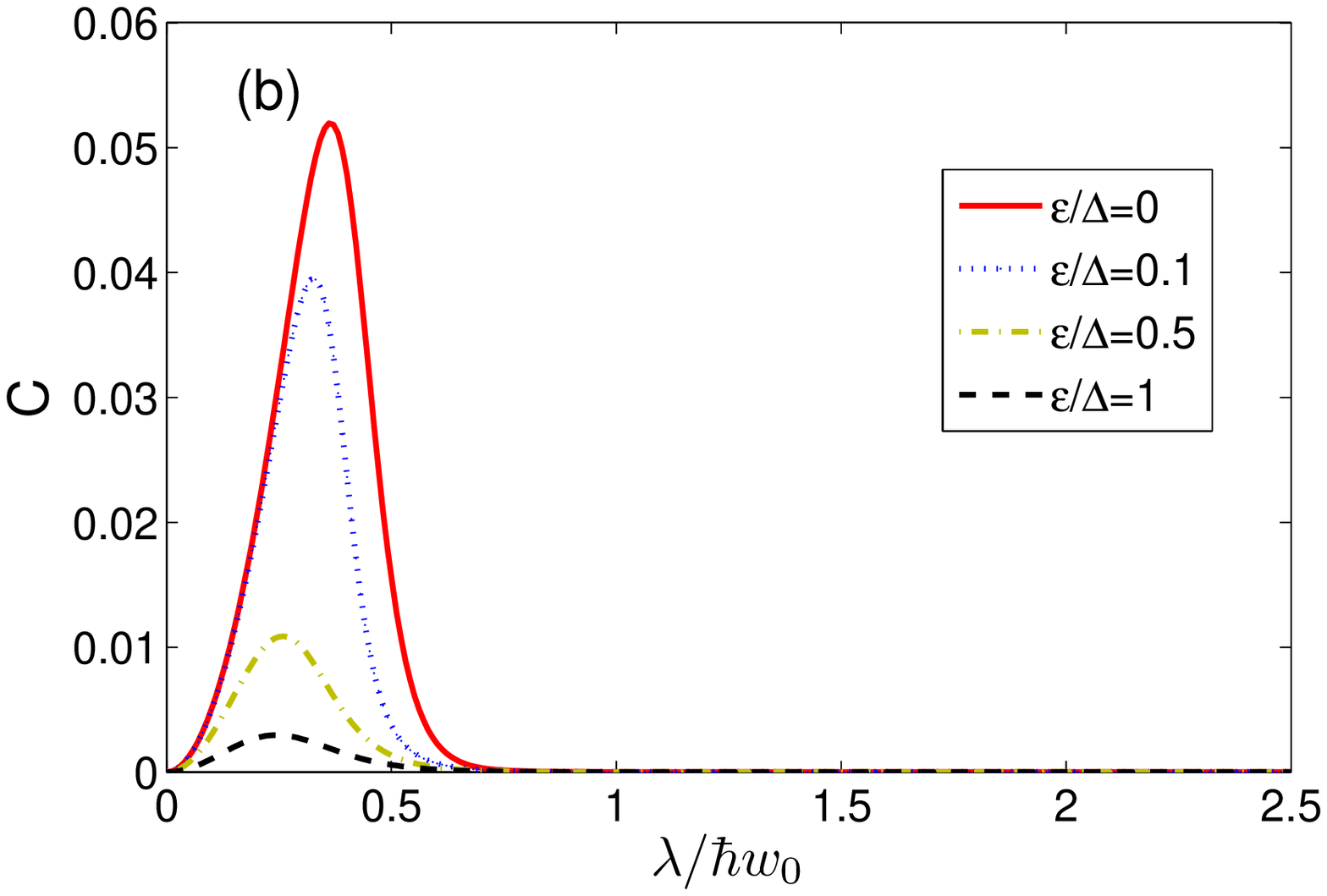}\\
  \includegraphics[width=0.9\columnwidth]{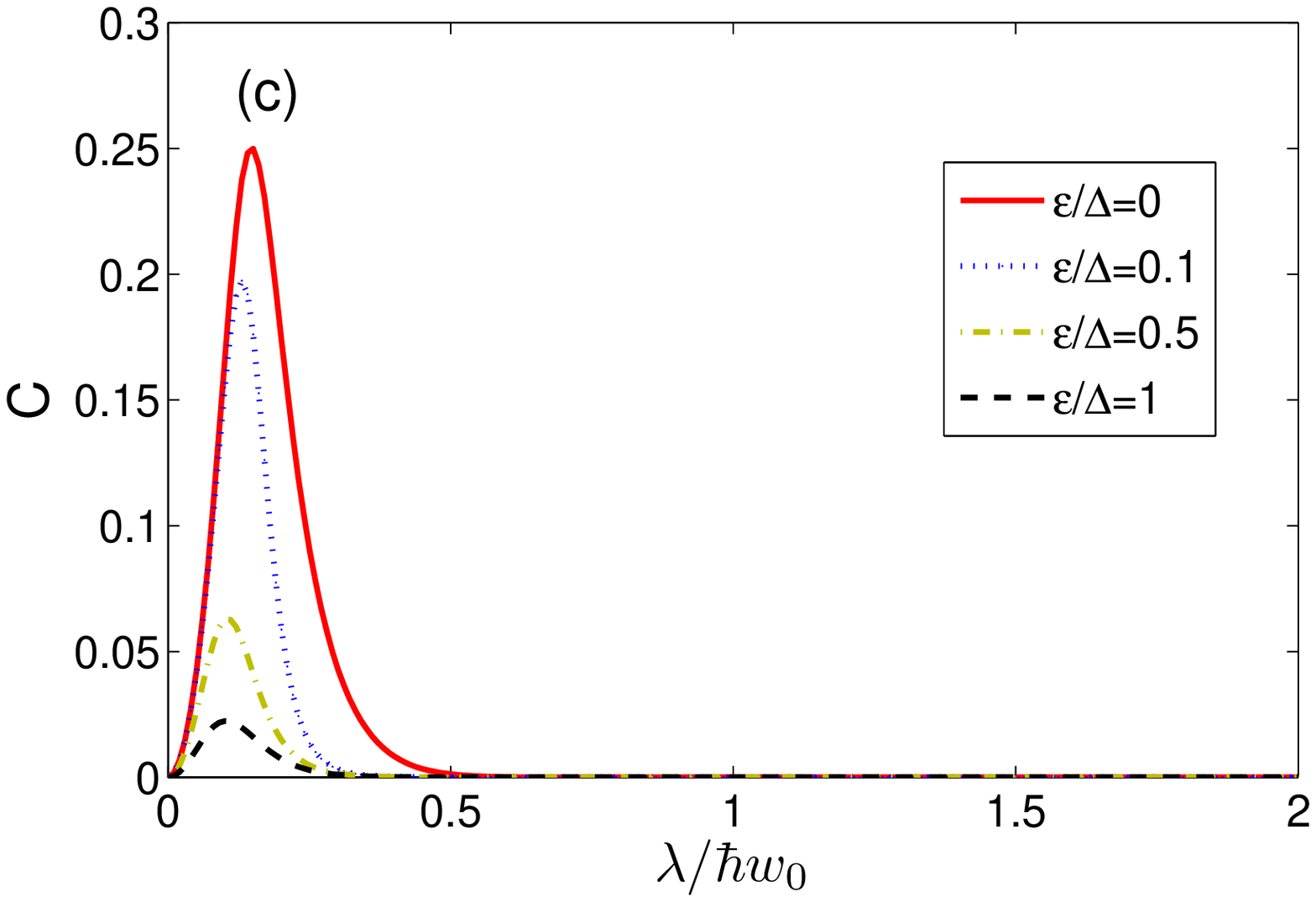} \caption{(Color
  online) The concurrence $C$ for the qubits 2 and 3 versus $\lambda/(\hbar w_0)$:
  (a) $\hbar w_0/\Delta=0.1$, (b) $\hbar
w_0/\Delta=1$, (c) $\hbar w_0/\Delta=10$.
  }\label{Fig.9.}
\end{figure}

To investigate the entanglement between any two qubits in the ground
state clearly, we use the Wootters's concurrence $C$
\cite{PRL-80-2245-1998} to quantify the entanglement between any two
qubits. The concurrence is defined as $C$ $=$ $\max\{$ $0$,
$\sqrt{e_1}-\sqrt{e_2}-\sqrt{e_3}-\sqrt{e_4}$ $\}$, where $e_1$,
$e_2$, $e_3$, and $e_4$ are four eigenvalues arranged in decreasing
order of the auxiliary matrix $\xi$ $=$ $\rho_{q_{23}}$
$(\sigma_y\otimes\sigma_y)$ $\rho_{q_{23}}^{*}$
$(\sigma_y\otimes\sigma_y)$ (We assume $\rho_{q_{23}}$ $=$
$Tr_{1}\{\rho_q\}$ is the reduced density matrix for the qubits 2
and 3) and $\sigma_y$ is the corresponding Pauli matrix. The
concurrence for the qubits 2 and 3 versus the qubit-oscillator
coupling strength is plotted in Fig. 9, which shows: (i) for the
small $\lambda$, the concurrence increases with increasing $\lambda$
and reaches to a maximum but then drops rapidly to zero and never
increases again as $\lambda$ increases further, meaning the
entanglement between any two qubits vanishes if the qubit-oscillator
coupling is strong enough; (ii) for $\epsilon=0$, the maximum
attainable concurrence is the biggest in the regime $\hbar
w_0>>\Delta$; and (iii) for $\epsilon\neq0$, the maximum attainable
concurrence decreases with increasing $\epsilon$, demonstrating the
fragility of the qubit-qubit entanglement in the ground state.

The numerical results mentioned above demonstrate that the regime
$\hbar w_0<<E_q$ is most suitable for preparing the squeezed states
in the oscillator, and the opposite regime $E_q>>\hbar w_0$ is most
proper for generating the entangled states between three qubits and
the oscillator. The nonclassical properties of the ground state are
analyzed through different intuitive quantifiers, which are
demonstrated to be highly susceptible to the variations in the bias
parameter.

\section{Conclusion}

We have analytically explored two extreme situations in the system
with three qubits interacting with an oscillator by using the
approach of adiabatic approximation: one situation considers a
high-frequency oscillator and the other one considers three
high-frequency qubits. We also numerically calculate the energy
spectra and ground-state properties for various combinations of the
system parameters which strengthens the outcomes of our analytical
derivations based on the adiabatic approximation. The nonclassical
properties of the ground state are analyzed through different
intuitive quantifiers, which are demonstrated to be highly
susceptible to the variations in the bias parameter. Interestingly,
we observe the phase-transition-like behaviors in the regime where
each qubit's frequency is far larger than the oscillator's
frequency, and find that the qubit-qubit entanglement in the ground
state vanishes if the qubit-oscillator coupling strength is strong
enough, in which the entropy of three qubits could keep larger than
one. Different from the system of one oscillator with only one
qubit, the minimum qubit-oscillator coupling strength needed to
produce the Schr\"{o}dinger-cat state is $1.25\hbar w_0$, which is
just one half of that in the single-qubit case. The
phase-transition-like behavior is also found in the regime where
each qubit's frequency is far larger than the oscillator's
frequency.

\section{Acknowledgement}

We would like to thank Dr. S. Ashhab and Prof. J. Larson for useful
discussions. This work is supported by the Major State Basic
Research Development Program of China under Grant No. 2012CB921601,
the National Natural Science Foundation of China under Grant No.
11374054, No. 11305037, No. 11347114, and No. 11247283, the Natural
Science Foundation of Fujian Province under Grant No. 2013J01012,
and  the funds from Fuzhou University under Grant No. 022513, Grant
No. 022408, and Grant No. 600891.

%%%%three

\end{document}